\documentclass[%
 reprint,
superscriptaddress,
 amsmath,amssymb,
 aps,
floatfix,
]{revtex4-2}

\usepackage[normalem]{ulem}
\usepackage{siunitx}
\usepackage{xcolor}
\usepackage{graphicx}
\usepackage{dcolumn}
\usepackage{bm}

\begin{document}

\preprint{APS/123-QED}

\title{History-dependent discharge of compressed particle rafts}

\author{Mario Nabernik}
 \affiliation{Graz University of Technology}
\author{Gregor Plohl}%
\affiliation{Infinite Biotech}%
\author{Kathrin Schulte}%
\affiliation{University of Stuttgart}
\author{Carole Planchette}
\affiliation{Graz University of Technology}
\email{carole.planchette@tugraz.at}

\date{\today}

\begin{abstract}
While particle-laden interfaces play a central role in many natural and industrial processes, predicting their mechanical properties remains a major challenge. These systems combine granular characteristics conferred by particle-particle contacts with elastic behavior originating from capillary interactions, making them  very sensitive to their history. Using the relaxation of uniaxially compressed particle rafts through a local constriction as a model experiment, we demonstrate the existence of a reproducible and continuous aging process. Aging is observed for both front- and back-compressed rafts and is characterized by a progressive increase in particle mobility and raft deformability. Macroscopic changes are seen, for example, in the extent of relaxation and are correlated with flow modifications observed at the mesoscopic level among which are increased particle fluxes, broader shear zones and enhanced particle rearrangements. While aging can be attributed unambiguously to the constrained passage of the particles through a constriction, its microscopic origin remains hypothetical, the results suggesting that contact lines around the particles may evolve. Beyond providing new insight into the effects of raft history, the proposed constriction flow experiment offers a simple method to control and compare aging in different particulate assemblies.  \\

\end{abstract}

\maketitle


\section{\label{sec:intro}Introduction}
Particle-laden interfaces exhibit fascinating mechanical properties combining elastic and granular characteristics~\cite{protiere2017}. Despite their relevance in natural~\cite{trottet2025} and industrial processes, their mechanical response remains poorly understood, which still limits their predictive modeling and broader exploitation. 
For example, effective elastic moduli can be derived from the wavelength of wrinkles developing under uniaxial compression~\cite{vella2004}, similarly to elastic membranes or insoluble surfactant monolayers in a Langmuir trough~\cite{milner1989, cerda2002}. Quasi-static experiments were first used to estimate an effective Young modulus~\cite{vella2004}, while later dynamic measurements independently characterized both the stretching and bending responses of particle rafts~\cite{planchette2012}. The stretching modulus was found to be independent of particle size, whereas the bending stiffness (or Young modulus) scales at first order with the square of particle diameter.

However, alongside these elastic descriptions, several observations also pointed toward a pronounced granular character. In particular, collective effects observed in bidisperse particle rafts cannot be reproduced by simple averaging of the two particle contributions. Variations in the efficiency of particle-particle contacts to transmit stress or possible percolation of one particle size network have been proposed to explain these observations~\cite{petit2016, planchette2018}. 
The granular character of particle rafts was independently evidenced in several studies using monodisperse assemblies, including raft aspect-ratio effects~\cite{cicuta2009}, Janssen-like stress screening~\cite{saavedra2018}, and the collapse dynamics of isolated chains~\cite{taccoen2016}. Similarly, experiments with rolling liquid marbles could show the importance of granular friction within the curved shells~\cite{takai2026}.

More recently, uniaxially compressed particle rafts relaxing through a small opening were used as model systems to probe the role of force-chain networks in interfacial stress transmission. When the opening was placed in the movable barrier used for compression, full relaxation was observed, whereas only partial unjamming occurred when the opening was located on the opposite barrier~\cite{plohl2022}. These results were interpreted as a consequence of the orientation and branching of the force-chain network, leading either to keystone removal and avalanches or to stable arches screening the opening. 

Furthermore, the robustness of unjamming depends on the preparation history of the raft~\cite{planchette2025}. Three different preparation protocols have been used that differed only by the level of mixing imposed to the particles prior to raft formation, while keeping other relevant parameters such as packing fraction and compression level unchanged. Results showed that increasing mixing intensity reduces the resistance of the network to shear and elongational stresses.
Such history effects suggest that the mechanical response of particle rafts is not controlled solely by their instantaneous packing state. In contrast to dry granular materials, interactions between particles are mediated by the liquid interface and may therefore depend on additional microscopic variables related to particle wetting and interface deformation~\cite{fournier2002,danov2005,liu2018}. The nature of these variables and their possible role in the emergence of memory effects remain largely unexplored.

The present paper aims to understand whether previously reported history effects appear progressively  and if they can be produced by the relaxation of the raft itself. Building on the previously developed raft-relaxation experiment, we investigate if and how repeated raft relaxation cycles modify the mechanical response of the rafts. Representative videos of raft relaxation are provided in the supplemental material for clarity. Measurements are carried  both at macroscopic and mesoscopic scales in order to address the following questions: can the observed history effects be seen as a continuous aging process? Which observables are affected and how? Which process causes the progressive modification of the raft age?  Finally, can the observed trends provide insight into the microscopic origins of memory and history dependence in interfacial granular systems?


\section{\label{sec:level2}Materials and Methods}

\subsection{Used particles and liquid}

Hydrophobic glass beads were used as particles. The silanization procedure follows Plohl et al.~\cite{plohl2022}, yielding to a contact angle of $126 \pm \SI{10}{\degree}$ when measured on single particles placed at the apex of a pendant drop ($n = 50$). The particle diameter is $127 \pm \SI{5}{\micro\metre}$ and their density is $\rho_\mathrm{s} = \SI{2500}{\kilogram\per\cubic\metre}$. 

The liquid is fresh distilled water. Its viscosity and surface tension remain constant over all runs.

\subsection{Experimental set-up and protocol}
\begin{figure}[t]
\centering
\includegraphics[width=0.9\linewidth]{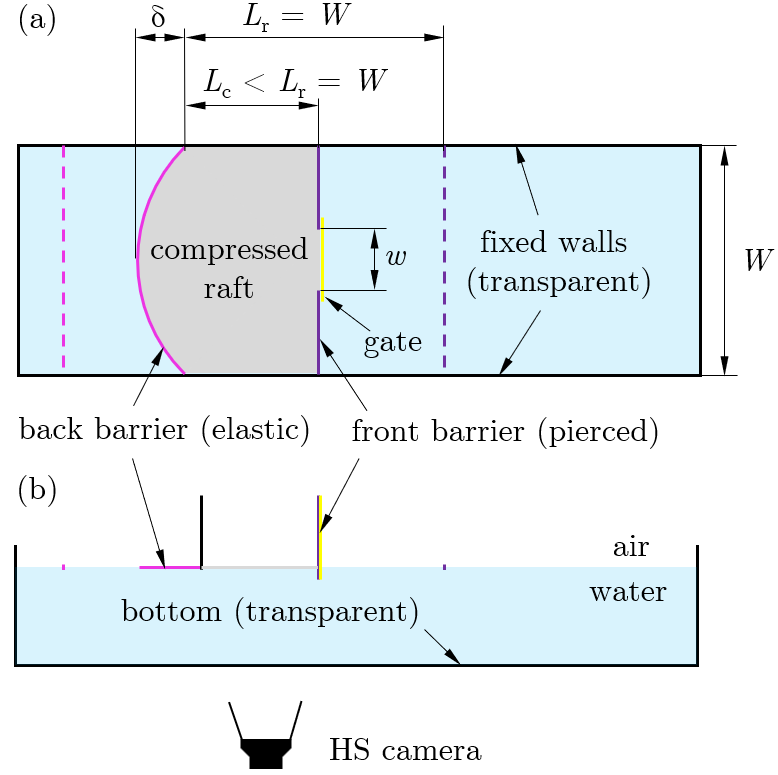}
\caption{Experimental setup from (a) bottom and (b) side. 
The trough of width $W$ confines particles between an elastic back barrier (pink) and a rigid front barrier (violet) pierced by an opening of width $w$, closed by a gate (yellow). The relaxed raft (length $L_\mathrm{r} = W$) is compressed to $L_\mathrm{c} < L_\mathrm{r}$ by moving either the back or the front barrier. The back barrier deflection $\delta$ provides the lineic pressure $\Pi$ acting on it.}
\label{fig:setup2}
\end{figure}

A transparent trough comprising two fixed lateral walls separated by $W= \SI{60}{\milli\metre} $, and two movable barriers at the front and back was used to compress particle rafts of $\displaystyle{K=\frac{L_{\mathrm{r}}-L_{\mathrm{c}}}{L_{\mathrm{r}}}}$ with $L_\mathrm{r}$ and  $L_\mathrm{c}$ the relaxed and compressed raft length, respectively. The set-up,  similar to the one used in previous studies~\cite{plohl2022, planchette2025}, is sketched in Fig.~\ref{fig:setup2}. The compression is either achieved by moving the back barrier or the front one. The back barrier is made of a calibrated rubber band whose deflection indicates the lineic pressure $\Pi$ developing at the back. The front barrier is a rigid wall pierced by an opening of width $w = \SI{10}{\milli\metre}$, initially closed by a gate. Opening the gate locally removes the constraint and creates a constriction through which the raft relaxes. 

Two cameras are used either separately or simultaneously to obtain the macroscopic evolution of the raft and additional mesoscopic details close to the constriction. In contrast to previous studies, both the relaxed raft length and its compression are kept constant with $L_\mathrm{r}=\SI{60}{\milli\metre}$ and $K=\qty{50}{\percent}$. The rafts are prepared as "tempered" rafts, meaning that the particles are not subjected to any mixing prior to raft formation, see also~\cite{planchette2025}. To do so,  the space between the barrier is set to approximately  $1.5 \, L_\mathrm{r}$, the particles are sprinkled  at the interface between them, and rearrange in a monolayer by gently blowing air with a pipette. The assembly is then compacted  by applying 5 quasi-static compression/decompression cycles until first wrinkles form/vanish providing a relaxed raft.
The distance between the barriers is then reduced from  $L_\mathrm{r}=\SI{60}{\milli\metre}$ to $L_\mathrm{c}=\SI{30}{\milli\metre}$ with a precision of \SI{1}{\milli\metre}. Note that this compression is made either by moving the front or the back barrier but never both. 
At this stage, the raft is ready to be studied.

Practically, the gate is opened and the cameras record the relaxation process. At the end, the particles that escaped from the confined area are then collected with the help of the movable barriers which are gently lifted and  translated back to their initial position with a spacing of approximately $1.5 \, L_\mathrm{r}$. The assembly is then compacted again and compressed using the same barrier, i.e. front or back, as before. The gate is opened again and the next relaxation recorded. This procedure is typically repeated 20 consecutive times, also called runs, with $n$ denoting the run number within a series.

More precisely, one series of 20 runs is studied for back compression with corresponding results discussed in Sec.~\ref{sec:back}. Unless stated otherwise, front compression was used representing three independent series of 20 to 30 runs each.

\subsection{Macroscopic variables}

\subsubsection{Unjammed and escaped areas}

\begin{figure}[h]
\centering
\includegraphics[width=0.95\linewidth]{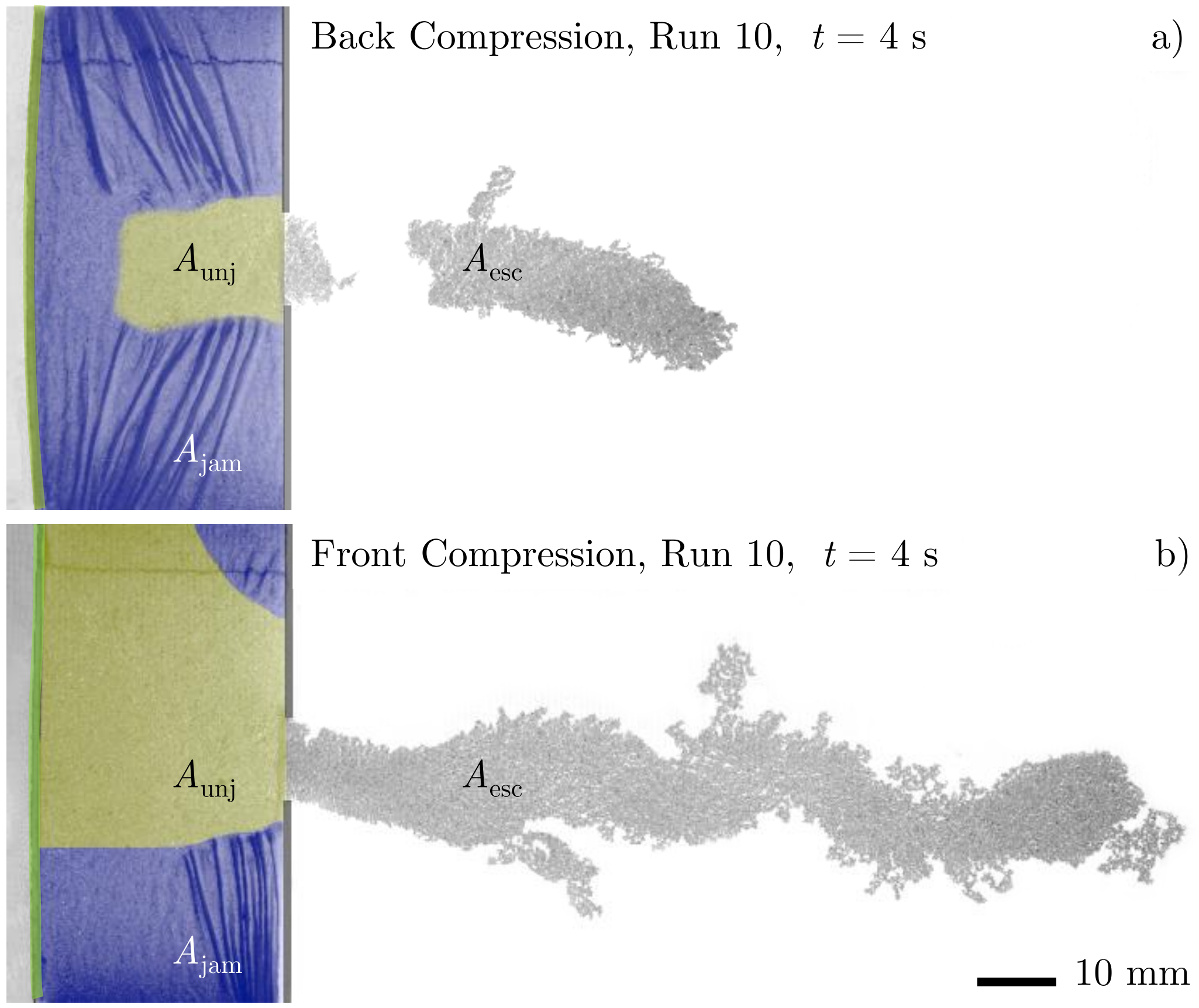}
\caption{Bottom view of relaxation images at $t=\SI{4}{\second}$ for run 10 and (a) back / (b) front compression. On the confined domain, the unjammed area is shown in yellow while the jammed one is colored in blue. The escaped particle assembly  is visible in gray outside the confined domain.}
\label{fig:illustrative}
\end{figure}

One camera always provides a general view of the relaxation from which we obtain $A_\mathrm{esc}$  the area initially free of particles which gets covered by the particles escaping the confined area and $A_\mathrm{unj}$, the initially jammed area that unjams. They are respectively indicated in gray and yellow in Fig.~\ref{fig:illustrative}. $A_\mathrm{esc}$ is obtained using a thresholding function. In these macroscopic videos, Voids larger than a single pixel, i.e \qty[parse-numbers=false]{200^2}{\micro\meter\squared} are resolved and excluded from the escaped area, whereas smaller voids are counted as particle-covered area, which may slightly overestimate $A_{\mathrm{esc}}$. It is then normalized by the area excess particle would cover providing $\displaystyle{A_\mathrm{esc}^*=\frac{A_\mathrm{esc}}{W\,(L_\mathrm{r}-L_\mathrm{c})}}$.  $A_\mathrm{unj}$ is derived from a machine learning approach using the Trainable Weka Segmentation plug-in  of IMAGEJ~\cite{arganda2017} and is normalized by the confined area giving $\displaystyle{A_\mathrm{unj}^*=\frac{A_\mathrm{unj}}{W \, L_\mathrm{c}}}$. For both the escaped and unjammed areas, the final values are further indicated by the subscript $\mathrm{f}$.


\subsubsection{Lineic pressure at the back}
The deflection $\delta$ of the rubber placed at the back of the trough  is measured in its center relative to its relaxed state. The initial value $\delta_\mathrm{i}$ corresponds to the state prior gate opening, while the final value $\delta_\mathrm{f}$ is taken once relaxation has ceased.  This measurement is used to follow the stress evolution over  consecutive runs. In the limit of small deflections, the pressure being directly proportional to the deflection ~\cite{planchette2025}, we obtain the normalized initial/final pressure as $\displaystyle{\frac{\Pi_{\mathrm{i/f}}(n)}{\Pi_{\mathrm{i}}(1)} \approx \frac{\delta_{\mathrm{i/f}}^{(n)}}{\delta_{\mathrm{i}}^{(1)}}}$.

\subsubsection{\label{sec:wrinkles}Wrinkle wavelength in the compressed raft}

\begin{figure}[h]
\centering
\includegraphics[width=0.99\linewidth]{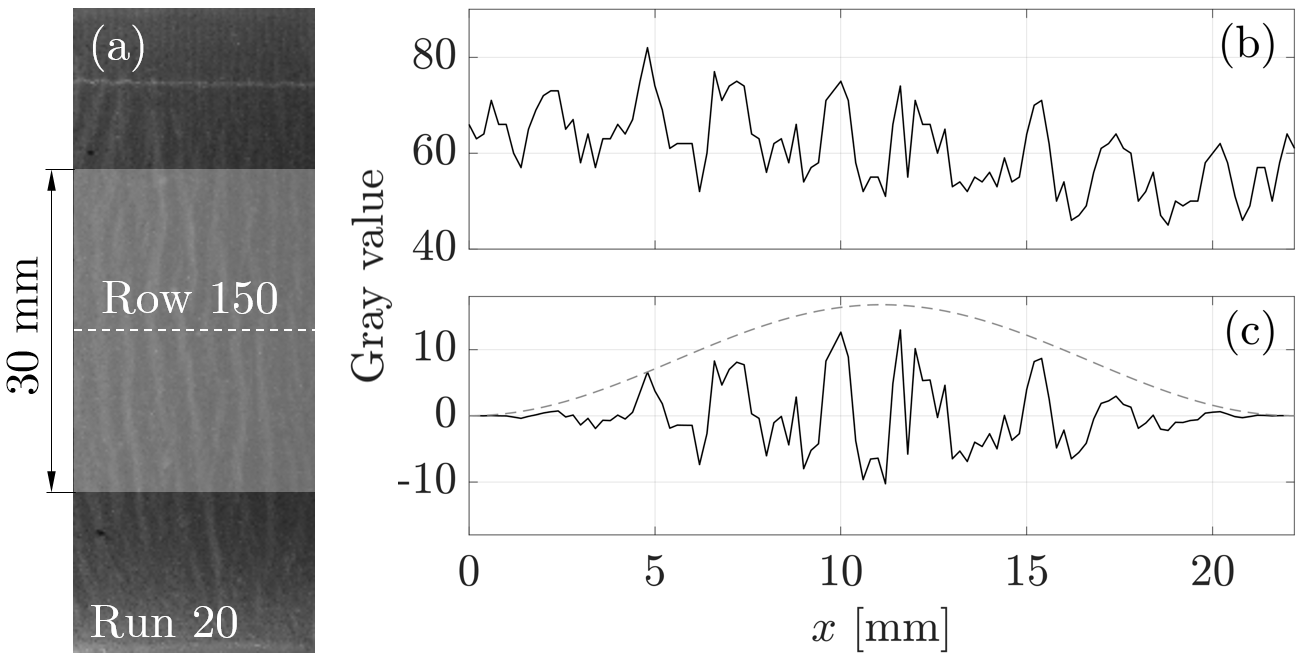}
\caption{Wrinkle spectral analysis procedure. (a) Light gray indicates the central half of the image used for the analysis  (rows 76--225, dashed line at row 150). (b) Raw gray value profile along the compression direction for row 150. (c) Corresponding detrended, mean-subtracted signal after application of the Hann window (dashed gray curve).}\label{fig:fig_lambda_method}
\end{figure}

Upon uniaxial compression, wrinkles form in the raft perpendicular to the compression direction. These are imaged with the low resolution camera keeping illumination and working distance unchanged, see Fig.~\ref{fig:fig_lambda_method} for illustration. The wrinkle wavelength is extracted from images of the compressed raft taken within the confined area. To reduce wall effects, only the central half of the image is considered as shown in Fig.~\ref{fig:fig_lambda_method}\,a in light gray. For each pixel row in this region, the gray value profile along the compression direction is extracted, see example of row 150 in Fig.~\ref{fig:fig_lambda_method}\,b. Each profile is then detrended, its mean is subtracted, and a Hann window is applied (Fig.~\ref{fig:fig_lambda_method}\,c) prior to computing the Fast Fourier Transform (FFT). The resulting spectra are averaged over all considered rows to yield the final power spectral density $P(\lambda)$.

The obtained spectra $P(\lambda)$ are then analyzed to characterize the evolution of the wrinkles over successive runs. 
For each spectrum, the maximum power density $P_\mathrm{max}$ and its respective wavelength $\lambda_\mathrm{max}$ is identified. The characteristic spatial scales $\lambda_\mathrm{small}$  and $\lambda_\mathrm{large}$ are then defined as the values corresponding to $P=0.75\,P_\mathrm{max}$ on both sides of the greatest peak.

\subsection{Flow kinematics within the constriction}

To follow the relaxation dynamic at mesoscopic level,  high speed videos focused around  the  constriction are analyzed. These videos are recorded with temporal and spatial resolution of 1000\,fps and \SI{29}{\micro\metre} per pixel. 

\begin{figure}
    \centering
    \includegraphics[width=0.7\linewidth]{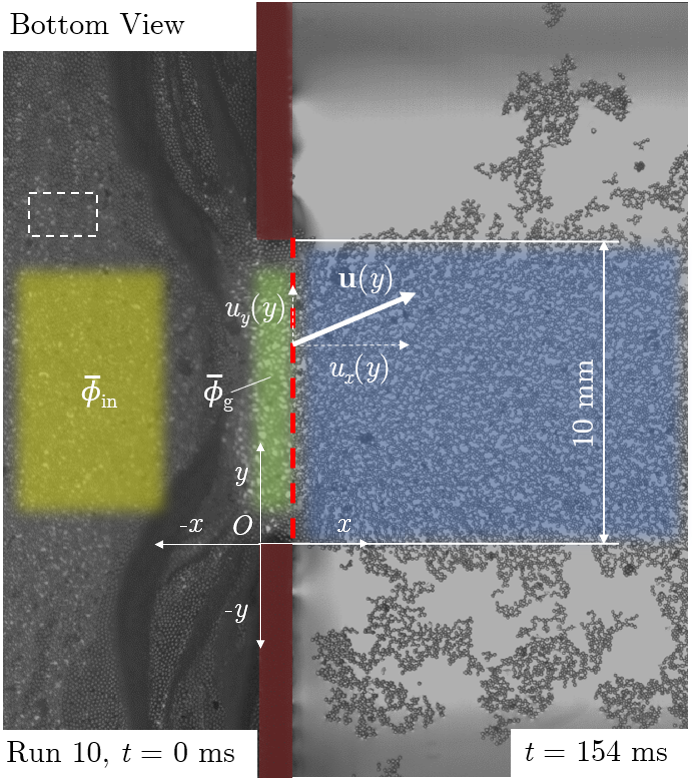}
    \caption{Bottom view with the constriction region (red dashed line and green area), part of the confined domain (yellow area), and part of the  domain occupied by escaped particles (blue area). Two time instants are combined to show the persistence of folds within the confined region at early times and the formation of an escaped particle assembly at later times. The dashed rectangle is used to follow individual particle rearrangements, see Sec.~\ref{sec:microdynamics}.}
    \label{fig:meso_variables}
\end{figure}

More precisely, this region of interest is further divided in three sub-domains colored in Fig.~\ref{fig:meso_variables}. The first sub-domain is the constriction itself (red dashed line and green area) from which the following quantities are calculated.

\subsubsection{Mean velocity, particle packing and particle flux}\label{sec:piv_gate}

The MATLAB-based Particle Image Velocimetry application \textsc{PIVlab}~\cite{thielicke2014, thielicke2021} is used to obtain the velocity field. Using a temporal sampling of \SI{500}{fps}, a multipass FFT window  deformation algorithm is employed with three successive passes of $128 \times 128$\,px (\SI{50}{\percent} overlap),  $64 \times 64$\,px and $32 \times 32$\,px (\SI{50}{\percent} overlap each) with a Gauss $2\times3$-point sub-pixel estimator with high correlation robustness.
The velocity field along the red dashed line of Fig.~\ref{fig:meso_variables} is then projected along the compression direction $x$. Spatial averaging along this line  provides the mean velocity $\overline{u}_x(t)$ at the constriction. This procedure is applied over the time interval ranging from 0 to \SI{4}{s}, providing its temporal evolution over the entire raft relaxation. In order to reduce short-time fluctuations, $\overline{u}_x(t)$ is smoothed using a sliding average over 100 frames, corresponding to \SI{200}{ms}.
The same analysis is repeated for each run number $n$.

In addition, an estimation of the mean particle packing at the constriction is deduced from applying a fixed gray thresholding function to the domain depicted in green in Fig.~\ref{fig:meso_variables}.  The spatially averaged particle surface fraction $\overline{\phi}_\mathrm{g}$ is calculated for each image as the ratio of the area covered by particles to the total analyzed area.   No temporal smoothing is applied to $\overline{\phi}_\mathrm{g}$ whose evolution can be followed as a function of time $t$ for each run number $n$.

Combining both quantities, the effective particle flux through the constriction is estimated as $\displaystyle{\overline{Q}_\mathrm{g}(t, n)=\overline{u}_x(t, n)\, \overline{\phi}_\mathrm{g}(t, n)}$. The accuracy of this PIV-based flux estimate is assessed by comparison with independent flux measurements obtained upstream of the constriction. Details of this validation procedure are provided in Appendix~\ref{app:subsec2}.

\subsubsection{\label{sec:initialVel}Instantaneous initial velocity profile} \label{sec:velocity_profile}

The early velocity field extracted along the red dashed line, see Sec.~\ref{sec:piv_gate},  is further used to determine the velocity profile across the constriction. For this analysis, a shorter time interval of duration  $\Delta  t=\SI{100}{\milli\second}$, centered around $t_\mathrm{i}=\SI{100}{\milli\second}$ is considered. The choice of $t_\mathrm{i}$  allows to focus on the early relaxation stage, just after  the main folds initially  found behind the constriction have disappeared. The temporal window of \SI{100}{\milli\second} is used to temporally average the profiles and reduce short-time fluctuations. The resulting profiles are plotted as a function of the transverse position $y$ along the red dashed line, taking the lower as origin, see Fig.~\ref{fig:meso_variables}. A finite edge velocity $u_{\mathrm{edge}}$ is obtained from the velocities measured approximately one particle diameter from each edge of the
constriction, $\displaystyle{\overline{u}_{\mathrm{edge}} = \displaystyle{\frac{1}{2}}\,\left[ u_x(y=\SI{0.15}{\milli\metre}) + u_x(y=\SI{9.85}{\milli\metre}) \right]}$. The individual values $\displaystyle{u_x (y=\SI{0.15}{\milli\metre})}$ and $\displaystyle{u_x (y=\SI{9.85}{\milli\metre})}$ are used to define lower and upper bounds of $\displaystyle{\overline{u}_\mathrm{edge}}$.
 
 To characterize the profile shape away from the walls, the profile measured in the region $\displaystyle{\SI{2}{\milli\metre}<y<\SI{8}{\milli\metre}}$ is fitted with a parabolic profile of the form 
 $\displaystyle{u_{x, \mathrm{theo}}= \overline{u}_\mathrm{edge}+a \, \frac{y}{w} \, \left( 1-\frac{y}{w} \right)}$ where $w$ denotes the width of the constriction and $a$ is an adjustable fitting parameter in m/s characterizing the profile curvature. This procedure is repeated for all run numbers $n$. 

\subsection{Raft rearrangements and deformation}

\subsubsection{Instantaneous initial shear rate maps}

To characterize the deformation of the relaxing raft, shear-rate maps are computed over the field of view shown in Fig.~\ref{fig:meso_variables}, using the same PIV settings as described in Sec.~\ref{sec:piv_gate}. The particle-free regions located on either side of the blue domain are excluded from the analysis to avoid artificial interpolation of the velocity field in areas where no particles are available for tracking. For each run, the mean velocity field is first computed by averaging the frames recorded over the same temporal window as for the velocity profile, i.e. from  $t_\mathrm{i}-\Delta t/2 =\SI{50}{\milli\second}$ to $t_\mathrm{i}+\Delta t/2 =\SI{150}{\milli\second}$. Here again, this choice is motivated by the need to eliminate early transient flow caused by the folds. The shear rate is then derived from the spatial gradients of this mean velocity field.

\subsubsection{Void statistics in escaped particle assembly} \label{sec:voids}
This analysis is performed on the blue domain shown in Fig.~\ref{fig:meso_variables}. To select a representative flow regime, both the initial transient associated with fold collapse and the late discharge that may comprise intermittent flows are excluded. Thus, this analysis is started at $t_\mathrm{i} = \SI{0.1}{\second}$ and carried out for a fixed duration of \SI{2}{\second}.

All frames recorded during this interval are analyzed with a fixed gray level threshold to identify the particles.  Only voids spanning at least 8\,px in area, corresponding to approximately \SI{6700}{\micro\metre\squared} are considered. This area  represents about half of the projected area of a single particle, \SI{12700}{\micro\metre\squared}.  This limit is set to filter out the central bright regions of the individual particles while retaining larger interstitial voids  developing within the escaped assembly. Individual voids are subsequently identified and characterized using the Analyze Particles function of ImageJ. For each run, each picture is analyzed and provides the void number density ${N}_\mathrm{v}(t)$ and the spatially averaged void area $\overline{A}_\mathrm{v}(t)$ for the corresponding instant $t$.

For each run $n$, the minimum and maximum values of ${N}_\mathrm{v}(t)$ and $\overline{A}_\mathrm{v}(t)$ for $\displaystyle{t_\mathrm{i}<t<t_\mathrm{f}}$ are identified and used to bound the temporally averaged quantities given by $\displaystyle{\left\langle {N}_\mathrm{v}\right\rangle= \frac{1}{t_\mathrm{f}-t_\mathrm{i}} \int_{t_\mathrm{i}}^{t_\mathrm{f}}{N}_\mathrm{v}(t)\,\mathrm{d}t}$
and $\displaystyle{\left\langle \overline{A}_\mathrm{v}\right\rangle= \frac{1}{t_\mathrm{f}-t_\mathrm{i}} \int_{t_{\mathrm{i}}}^{t_{\mathrm{f}}}\overline{A}_\mathrm{v}(t)\,\mathrm{d}t}$,
respectively.

\subsubsection{\label{sec:microdynamics} Individual particle dynamics in the confined domain}

To document elementary particle rearrangements, a small region of interest  located in the high-shear, converging-flow region upstream of the constriction is selected. This region is marked by the dashed rectangle in Fig.~\ref{fig:meso_variables} and imaged with a spatial resolution of \SI{3.2}{\micro\metre} per pixel enabling individual particle tracking. Using the ImageJ plugin TrackMate~\cite{ershov2022, tinevez2017}, $\mathbf{r}_i(t)$, the position of each particle $i$ is followed as a function of time $t$.
For each particle $i$ and increment $\Delta t = 5$~ms, the non-affine
displacement $D^2_{\min}$ of Falk and Langer~\cite{falk1998} is computed
as the residual of the best affine mapping $\mathbf{J}_i$ of the local
neighborhood between $t$ and $t+\Delta t$:
\begin{equation*}
\begin{split}
D^2_{\min,i}(t) = \min_{\mathbf{J}_i} \sum_{j \in \mathcal{N}_i}
\Big\| \big[\mathbf{r}_j(t+\Delta t) - \mathbf{r}_i(t+\Delta t)\big] \\
{}- \mathbf{J}_i \big[\mathbf{r}_j(t) - \mathbf{r}_i(t)\big] \Big\|^2 ,
\end{split}
\end{equation*}
where $\mathcal{N}_i$ are the neighbors of $i$. A low $D^2_{\min}$
indicates near-affine motion indicating that the particle moves with its cage while a high value flags significant local rearrangement relative to the immediate neighbors. Particles are colored according to their $D^2_{\min}$ value and labeled  to further provide their individual trajectories over time.

\section{\label{sec:level3} Results and discussion}

\subsection{Relaxation behaviors}

\subsubsection{\label{sec:back}Back compression}

\begin{figure}[h]
        \centering
        \includegraphics[width=0.95\linewidth]{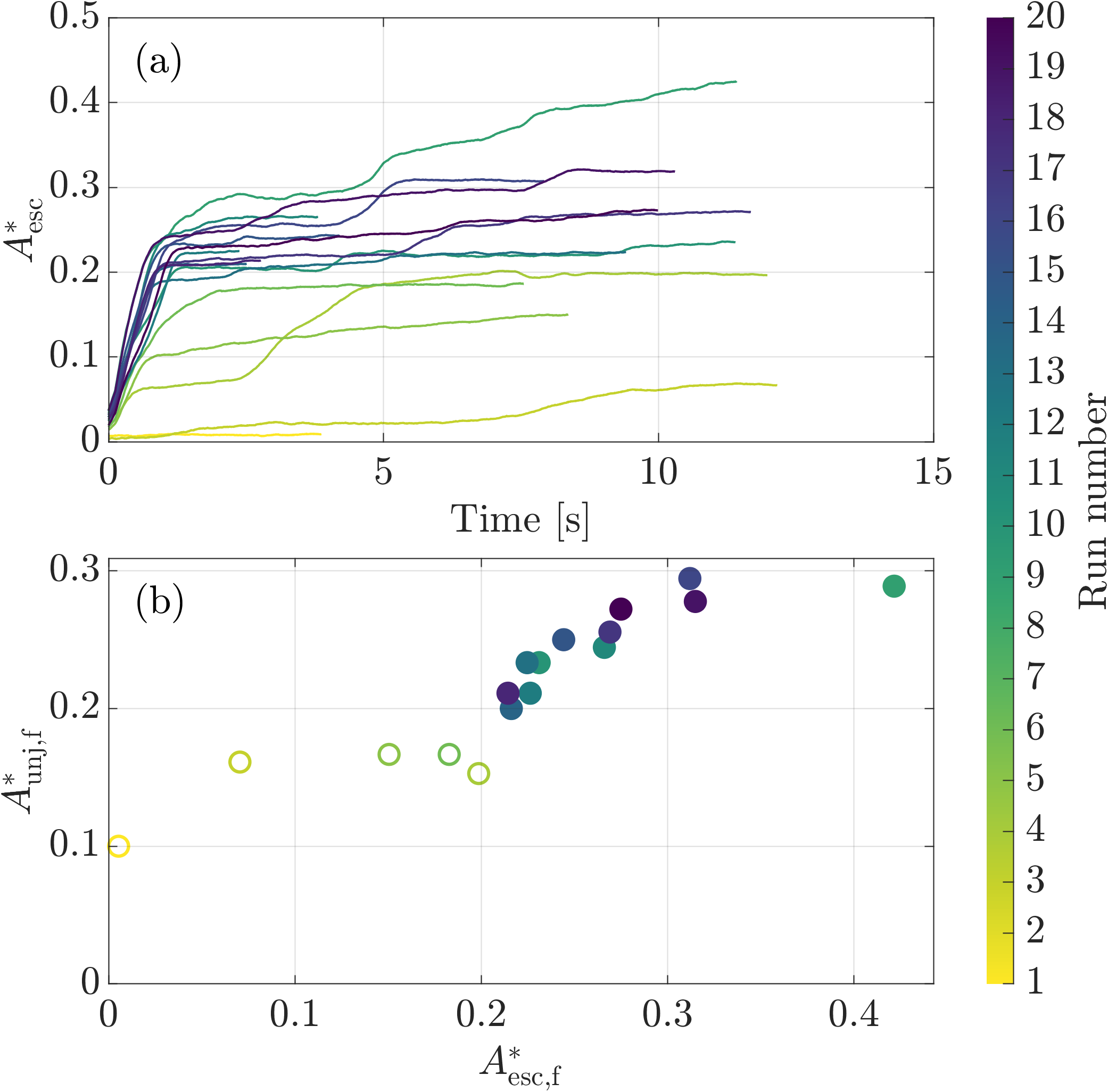} 
    \caption{Back compression. (a) Temporal evolution of normalized escaped area $A^*_\mathrm{esc}$ for 20 consecutive runs. (b) Final normalized unjammed area $A^*_\mathrm{unj,\,f}$ as a function of the final normalized escaped area $A^*_\mathrm{esc,\,f}$. Empty/filled symbols indicate unjamming stopping before/at the back rubber. Same color scale for run number $n$.}
    \label{fig:aesc_back}
\end{figure}


The evolution of the normalized escaped area obtained under back compression is reported as a function of time for different run numbers in Fig.~\ref{fig:aesc_back}\,a. Illustrative movies are provided in supplemental material. All curves exhibit an initial approximately linear increase followed by a plateau, possibly disturbed by discrete positive jumps. This behavior is consistent with previous observations of back-compressed rafts~\cite{plohl2022}. Interestingly, both the initial slope, which provides a measure of the escape kinetics, and the final plateau level, which characterizes the final degree of relaxation, increase rapidly during the first runs. Beyond approximately $n\approx 5$, and except for run 9, the differences become smaller. 


To go further, the final unjammed  is plotted as a function of the final escaped area in Fig.~\ref{fig:aesc_back}\,b. 
During the first five runs, the escaped area increases substantially while the unjammed one remains nearly constant. This behavior is consistent with incomplete unjamming marked by empty symbols and shown in Fig. \ref{fig:illustrative}\,a.  From run 6 onward, unjamming extends until the back of the raft forming a channel of almost constant width close to $w$. In this regime, the escaped and unjammed areas evolve almost proportionally. No clear systematic increase is observed and the slight variations are attributed to small changes in the width of the unjammed channel developing behind the opening.

Repeated cycling of back compressed rafts progressively enhances the raft relaxation, in agreement with  memory effects  reported for rafts prepared with increasing mixing intensity~\cite{planchette2025}. The weaker evolution observed at large run numbers may be attributed to the force-chain orientation associated with back compression~\cite{plohl2022}. As arching structures screen the stress acting on either side of the constriction, only particles located directly behind it actively participate in the raft relaxation. Since the escaped particles are recycled after each run, essentially the same region is repeatedly reprocessed, while the rest of the raft remains practically unchanged. This localized relaxation may explain the progressive saturation observed at large run numbers.

\subsubsection{Front compression}

\begin{figure}[h]
        \centering
        \includegraphics[width=0.95\linewidth]{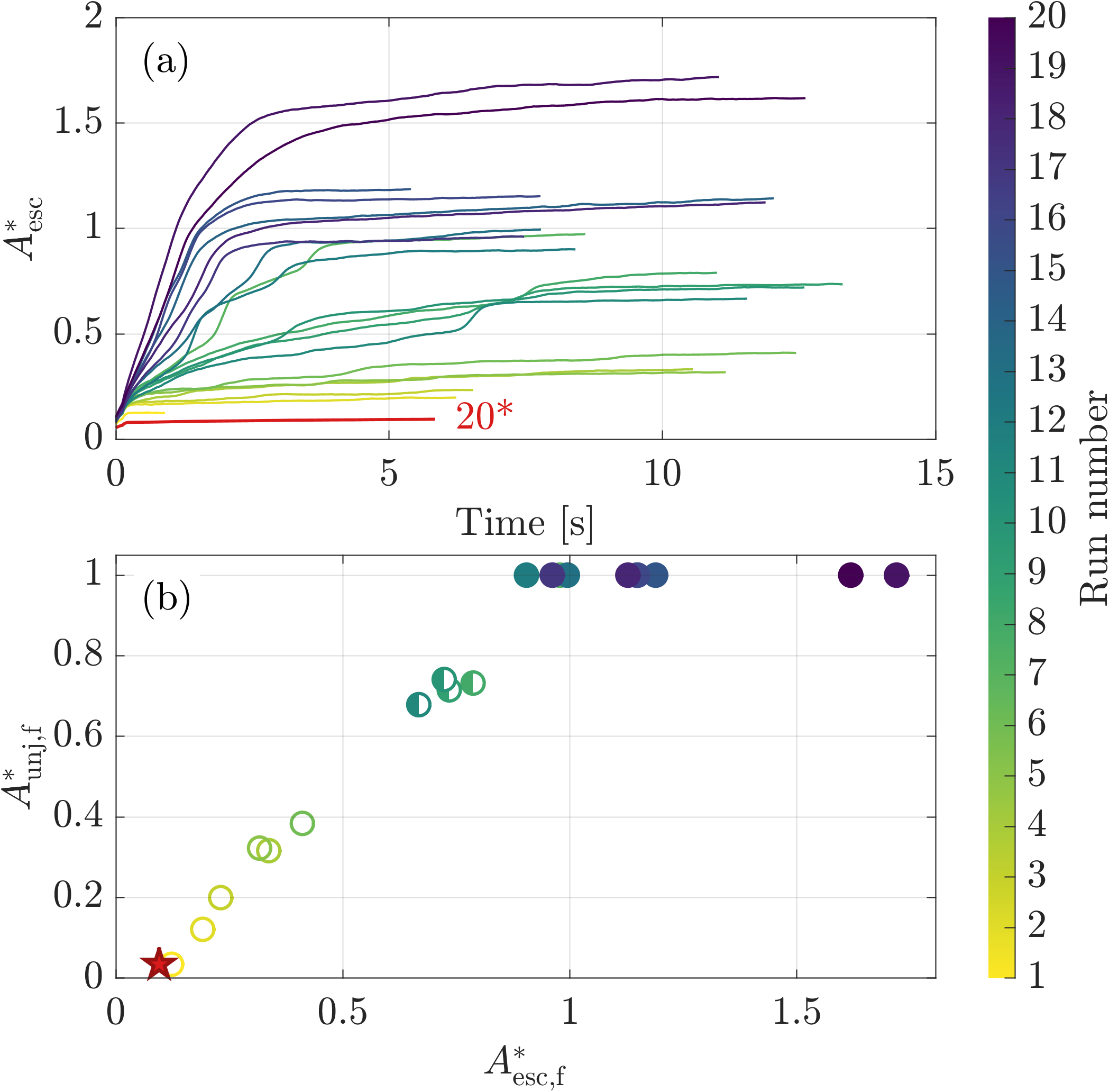}
    \caption{Front compression. (a) Temporal evolution of normalized escaped area $A^*_\mathrm{esc}$ for 20 consecutive runs.(b) Final normalized unjammed area $A^*_\mathrm{unj,\,f}$ as a function of the final normalized escaped area $A^*_\mathrm{esc,\,f}$. Empty/half-filled/filled symbols indicate the unjammed region reaching none/one/both lateral walls, respectively. Same color scale to indicate $n$, red line/symbol correspond to control experiment.}
    \label{fig:aesc_front}
\end{figure}

In contrast to back compressed rafts, front compressed rafts evidence a stronger or more persistent evolution over the investigated run number, see also movies in the supplementary material for direct comparison. As shown in Fig.~\ref{fig:aesc_front}\,a, both the relaxation kinetics and final escaped area increase continuously with run number, with no clear indication of saturation. 

The final unjammed and escaped areas are reported in Fig.~\ref{fig:aesc_front}\,b.  For small run numbers, only a limited unjammed region develops, which does not reach either lateral wall of the trough (empty symbols). For intermediate run numbers, the unjammed domain often extends to one of the two lateral walls as visible in Fig. \ref{fig:illustrative}\,b (half full symbols). In both regimes, the escaped and unjammed areas increase approximately proportionally, showing that the release of additional particles is accompanied by the relaxation of an increasing fraction of the raft. For large run numbers, the  unjammed region systematically connects both lateral walls (full symbols). Consequently, $A^*_\mathrm{unj,\,f}$ saturates at its maximum value of unity. $A^*_\mathrm{esc,\,f}$ continues to increase, indicating that particles are progressively dragged through the constriction beyond those required to simply relax the initial compression.

\begin{figure}[h]
\centering
\includegraphics[width=0.95\linewidth]{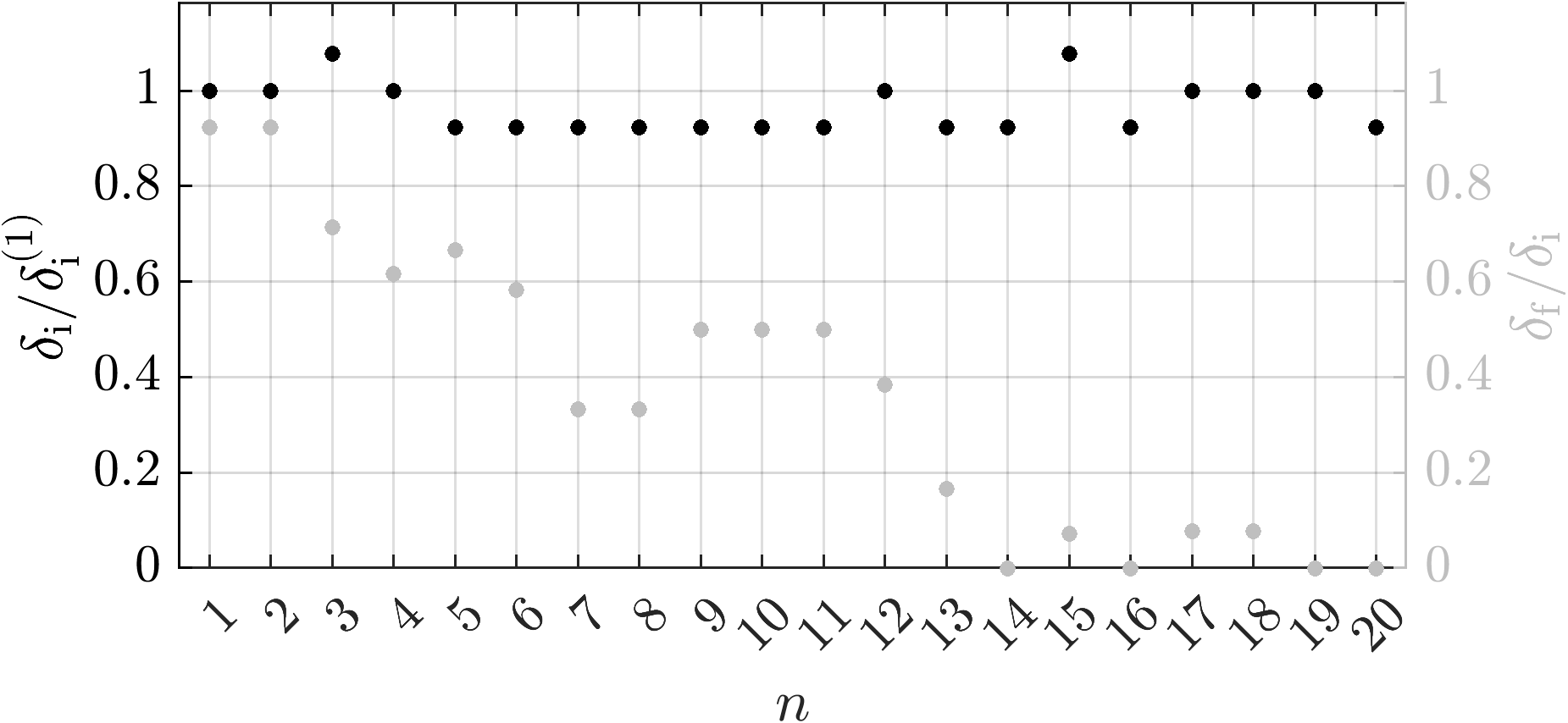}
\caption{Normalized rubber deflection as a function of run number for front compression. 
Black symbols (left axis): initial deflection $\delta_\mathrm{i}/\delta_\mathrm{i}^{(1)}$ confirming similar stress transmission. Gray symbols (right axis): final  deflection $\delta_\mathrm{f}/\delta_\mathrm{i}$ confirming a progressive increase in stress relaxation over run number.}
\label{fig:rubber}
\end{figure}

While a tentative explanation for the observed evolution of $A^*_\mathrm{unj}$ and $A^*_\mathrm{esc}$ is a gradual drift of the experimental conditions, it is indeed ruled out by the following analysis. The escaped particles are carefully collected after each run so that both the relaxed raft length and the amount of material remain essentially unchanged. Further, the same compression length is set with a precision of about \SI{1}{\milli\metre}. These identical compressed raft geometries produce nearly constant deflection of the back rubber barrier, indicating that the stress transmission through the raft  remains unchanged within experimental uncertainty, see Fig.~\ref{fig:rubber}. The observed evolution can therefore not be attributed to variations in particle number, compression level, or stress mobilization at the lateral walls. 

To further clarify this point, an additional control experiment was performed by repeatedly applying the 20 front compressions including intermediate cycling for better compaction but without opening the gate, thus preventing particles to flow through the constriction. Only after the $20^{\mathrm{th}}$ compression was the gate opened and the relaxation recorded, yielding a single set of observables. This run, hereafter denoted with 20*, carries the compression history of 20 cycles but the discharge history of a first run. In this case and as indicated by the red line and symbol in Fig.~\ref{fig:aesc_front} no comparable evolution was observed demonstrating that cyclic loading alone is insufficient to produce the reported aging effects.

Taken together, these observations strongly suggest that the observed aging  originates from modifications induced by the repeated relaxation process itself. We therefore use $n$ as a measure of the raft age in the following.
Although similar trends are observed for both compression configurations, back compression probes only a limited fraction of the raft and its relaxation seems to reach a plateau. The following sections therefore focus on front-compressed rafts, for which aging appears more pronounced over a  larger number of runs. 

\subsection{\label{sec:evolution}Evolution of flow through the constriction}

\subsubsection{Mean particle velocity, packing and flux through the constriction}

\begin{figure}[h]
\centering
\includegraphics[width=0.99\linewidth]{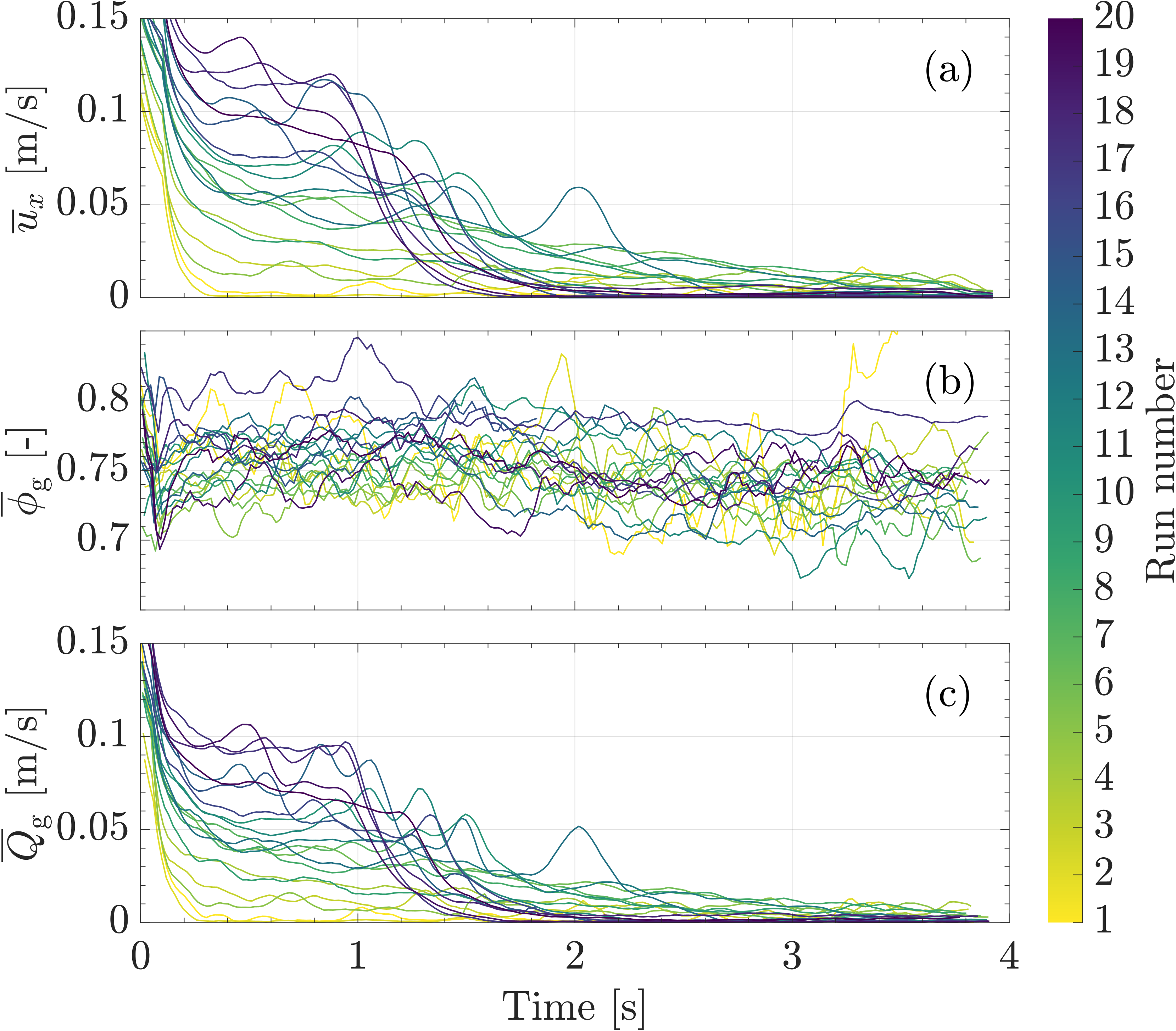}
\caption{Temporal evolution of (a) mean velocity component along the compression direction $\overline{u}_x$, (b) spatially averaged 
particle surface fraction $\overline{\phi}_\mathrm{g}$, and (c) effective particle 
flux $Q_\mathrm{g} = \overline{u}_x\,\overline{\phi}_\mathrm{g}$.  All these quantities are obtained at the constriction and plotted for 20 consecutive runs from yellow (run 1) to purple (run 20).}
\label{fig:mean_velocity}
\end{figure}

We now quantify the evolution occurring at the constriction by considering the mean particle velocity, mean particle fraction and corresponding particle flux measured over the entire relaxation process, see Fig.~\ref{fig:mean_velocity}\,a--c.

The mean velocity at the constriction,  $\overline{u}_x$,  increases continuously with run number. For a given run, the highest velocities are typically observed during the early stages of the discharge, where the effect of aging is very pronounced. At later times, large run numbers velocities may vanish earlier than intermediate ones because all particles have already escaped from the confined region. In contrast, no significant trend can be identified for the particle fraction $\overline{\phi}_\mathrm{g}$, which fluctuates with an amplitude of $\pm 0.05$ around a mean value of $0.75$ for all runs. As a result, the particle flux closely follows the evolution of the mean velocity, confirming that the increase in particle discharge is primarily driven by the increase in particle velocity rather than by changes in their local packing.

These results demonstrate that aging affects not only the macroscopic raft relaxation but also the flow through the constriction. To further characterize this evolution, we now examine the early velocity profiles developing within the constriction.

\subsubsection{Velocity profile}

\begin{figure}[h]
\centering

\hspace*{4.5mm} 
\includegraphics[width=0.99\linewidth]{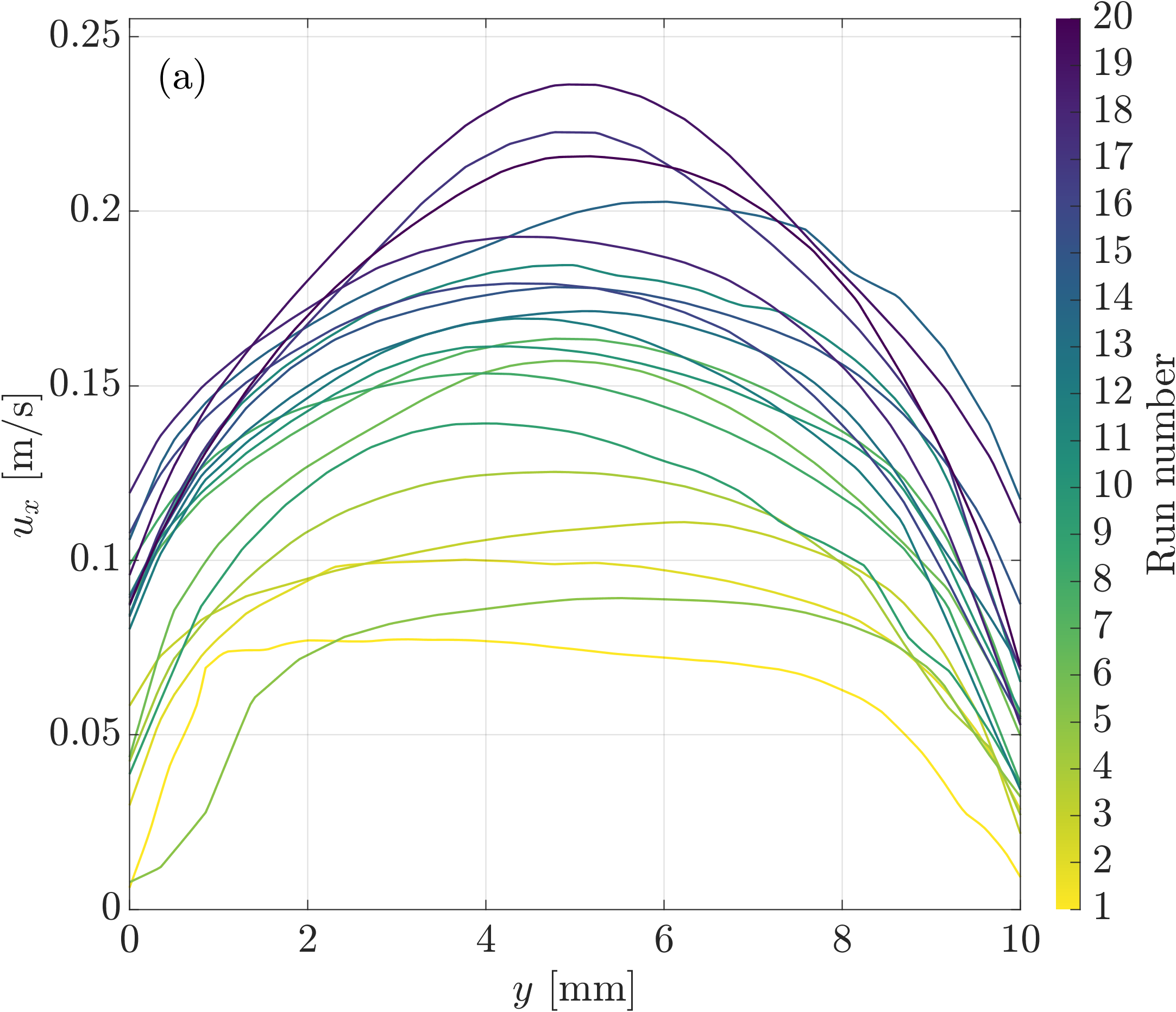}

\vspace{3mm}

\hspace*{0mm} 
\includegraphics[width=0.9\linewidth]{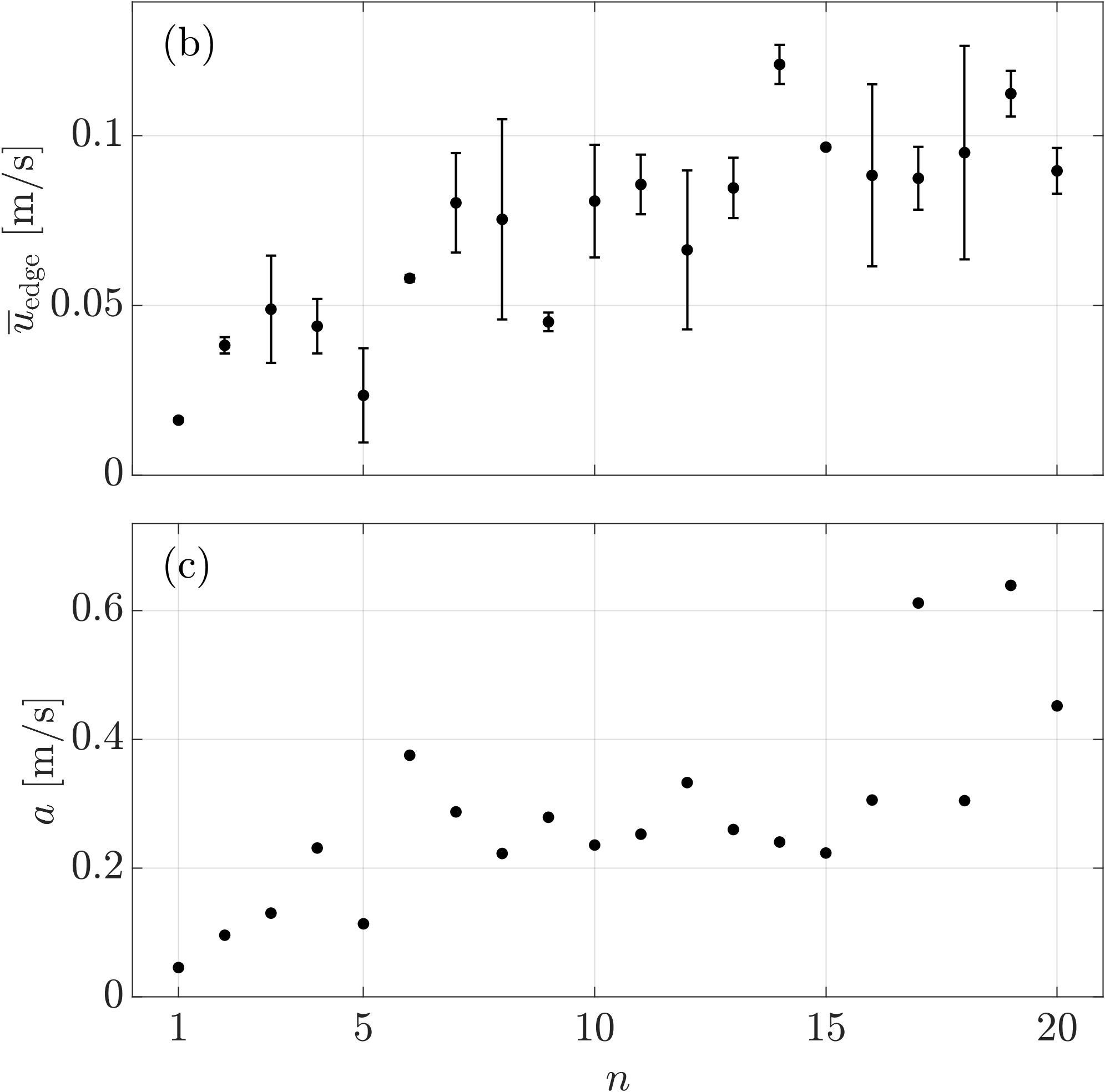}
\caption{(a) Velocity profile as a
function of transverse position across the constriction for all runs from yellow (run 1) to purple (run 20). (b) Wall velocity estimated as the mean of the velocities measured at a distance \SI{0.15}{\milli\metre} from the edges; with bars bounding the two values. (c) Parabolic fitting parameter $a$ (in m/s), obtained from a fit over the range $\displaystyle{\SI{2}{\milli\metre} < y < \SI{8}{\milli\metre}}$.}
\label{fig:velocity_profile}

\end{figure}

The velocity profiles obtained for 20 consecutive runs are plotted in Fig.~\ref{fig:velocity_profile}\,a. Comparing early to late runs, two main differences arise. First, the slip velocity at the wall  increases with run number shifting the entire profiles toward higher values. Second, the shape of the profile itself gets modified with rather flat, plug-like profiles at the beginning and parabolic ones, as classically observed for viscous Hagen-Poiseuille flows for later runs.

To better quantify this evolution, the velocity at the wall, defined as the velocity measured at $\SI{0.15}{\milli\metre}$ from the wall (see Sec.~\ref{sec:velocity_profile}), is reported in Fig.~\ref{fig:velocity_profile}\,b bounded by its maximum and minimum values as bars. A clear increase of $\overline{u}_\mathrm{edge}$ is observed from less than $\SI{0.02}{\metre/\second}$ for first run  to  values close $\SI{0.10}{\metre/\second}$ for larger run numbers. 

The evolution of the profile shape is followed by plotting the parameter $a$ obtained
by fitting the experimental profiles with  $\displaystyle{u_{x, \mathrm{theo}}= \overline{u}_\mathrm{edge}+a \, \frac{y}{w} \, \left( 1-\frac{y}{w} \right)}$ as a function of the run number, see Fig.~\ref{fig:velocity_profile}\,c. The parameter $a$ increases by more than a factor of five over the 20 runs, indicating a progressive increase in profile curvature and a departure from plug-like motion.
Consistently, the ratio between the mean and maximum values of $u_{x} - u_\mathrm{edge}$, i.e. of the velocity profile without wall slip, is found to be $0.67 \pm 0.02$  for $n>5$. This is consistent with the theoretical value  of $2/3$ given by a parabolic profile.

The evolution of the velocity profiles from a plug-like to a parabolic shape, together with the increase in wall slip, demonstrates that aging affects not only the mobility of the raft as a whole but also the relative mobility of the particles within it. The resulting increase in internal deformation indicates that particle rearrangements become progressively easier with repeated cycling. The larger wall-slip velocities suggest a reduced resistance to relative motion at the boundaries. These findings suggest that repeated cycling progressively modifies the dynamics of particle contacts, reducing both particle-particle and particle-wall resistance to relative motion.

\subsection{Deformation beyond the constriction}

\subsubsection{Shear rate maps}
\begin{figure*}[t]
\includegraphics[width=0.9\textwidth]{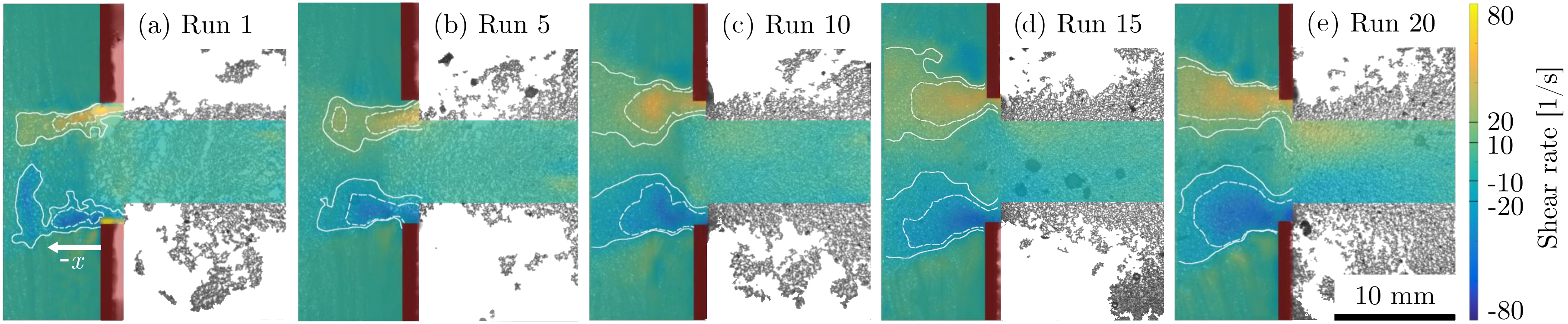}
\caption{(a--e) Shear rate fields for runs 1, 5, 10, 15 and 20; continuous and dashed contours mark the regions  exceeding \SI{10}{\per\second} and \SI{20}{\per\second} respectively. }
\label{fig:shear1}
\end{figure*}

\begin{figure}[h]
\centering
\hspace*{0mm} 
\includegraphics[width=0.9\linewidth, trim=30 20 35 35, clip]{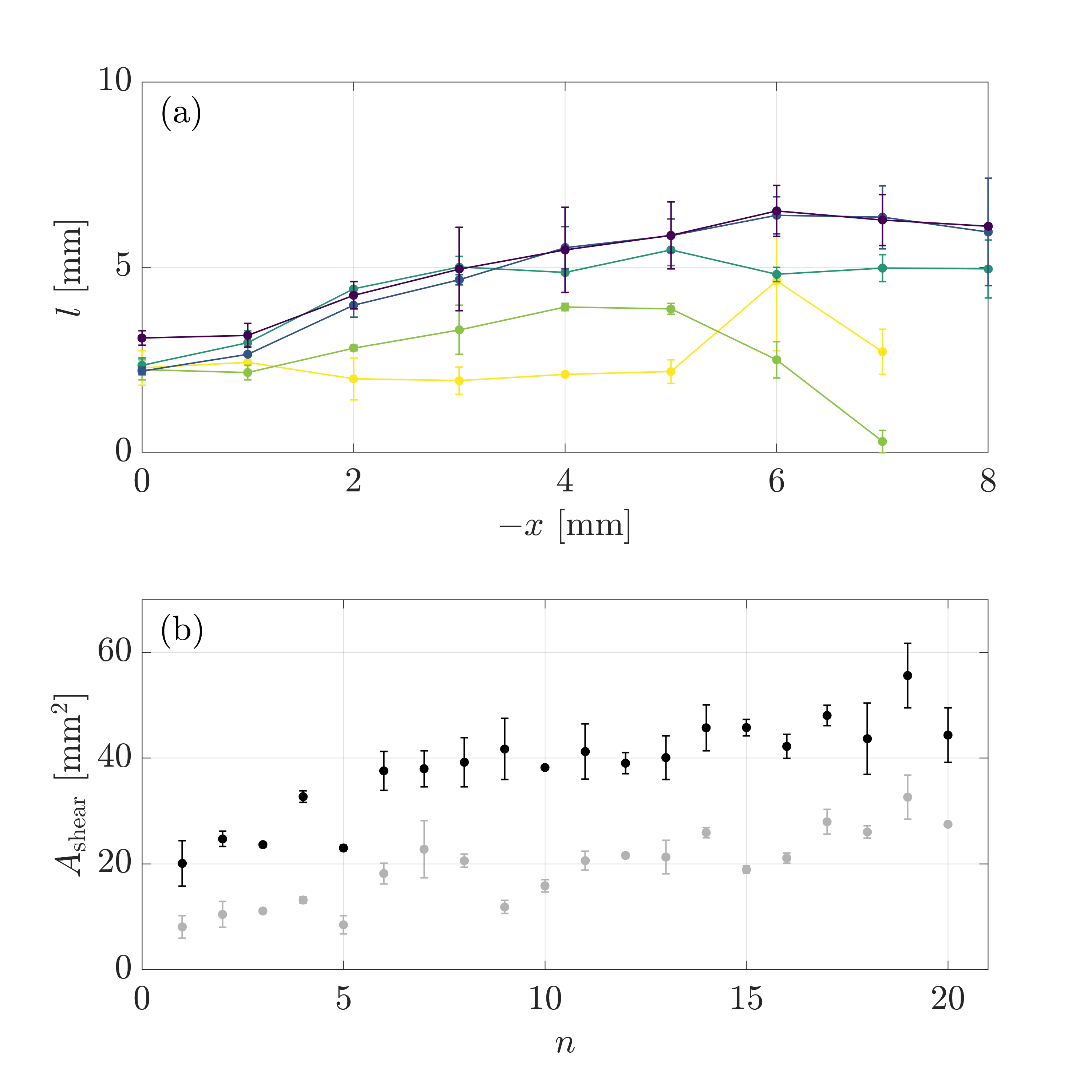}
\caption{ (a)~Width $l$ of the shear rate field zone exceeding \SI{10}{\per\second} as a function of position $x$ upstream of the constriction for runs 1, 5, 10, 15, and 20. (b)~Areas occupied by shear rates above  \SI{10}{\per\second} (black) and \SI{20}{\per\second} (gray) as a function of run number. In both panels, the marker indicates the mean of the two lateral measurements and the bars their spread.}
\label{fig:shear2}
\end{figure}

While the velocity profiles provide information on the deformation occurring at the constriction itself, they do not describe how this deformation is distributed throughout the raft. To address this question, shear-rate maps are now considered.
Illustrative maps obtained for runs 1, 5, 10, 15 and 20 are shown in Fig.~\ref{fig:shear1}\,a--e. The highest shear rates are observed on either side of the constriction within the confined region. 

Comparison of the maps reveals two main trends developing with increasing run number. First, the overall level of deformation increases with run number, as evidenced by the larger regions exceeding shear rates of \SI{10}{\per\second} and \SI{20}{\per\second} as indicated by the continuous and dashed contours, respectively. This observation is confirmed quantitatively in Fig.~\ref{fig:shear2}\,b, where these areas are plotted as a function of $n$ for all runs. Both increase continuously over the investigated range numbers without reaching a clear plateau.

Second, the shape of the sheared regions is modified. Initially limited to two rather narrow bands aligned with the compression direction, they progressively broaden toward conically shaped domains. To quantify this effect, the width $l$ of the regions enclosed by the \SI{10}{\per\second} and \SI{20}{\per\second} contours is plotted as a function of the upstream distance $-x$. The measurements obtained for runs 1, 5, 10, 15 and 20 confirm the progressive widening. The corresponding cone angle, estimated as $2 \tan^{-1}(\mathrm{d}l/2\mathrm{d}x)$, is approximately $55^{\circ}$ close to the  $60^{\circ}$ angle expected for force transmission along close hexagonal packing direction.

Together, these results indicate that aging progressively redistributes deformation within the raft. While shear is initially concentrated in  narrow bands at the edges of the constriction, it progressively spreads over larger and wider areas with increasing run number. Thus, although the stress measured at the back of the raft remains essentially unchanged, the deformation it causes  within the raft during relaxation is significantly modified by aging.  

\subsubsection{Evolution of the escaped particle assemblies}
\begin{figure}[h]
\centering

\hspace*{0mm} 
\includegraphics[width=0.9\linewidth]{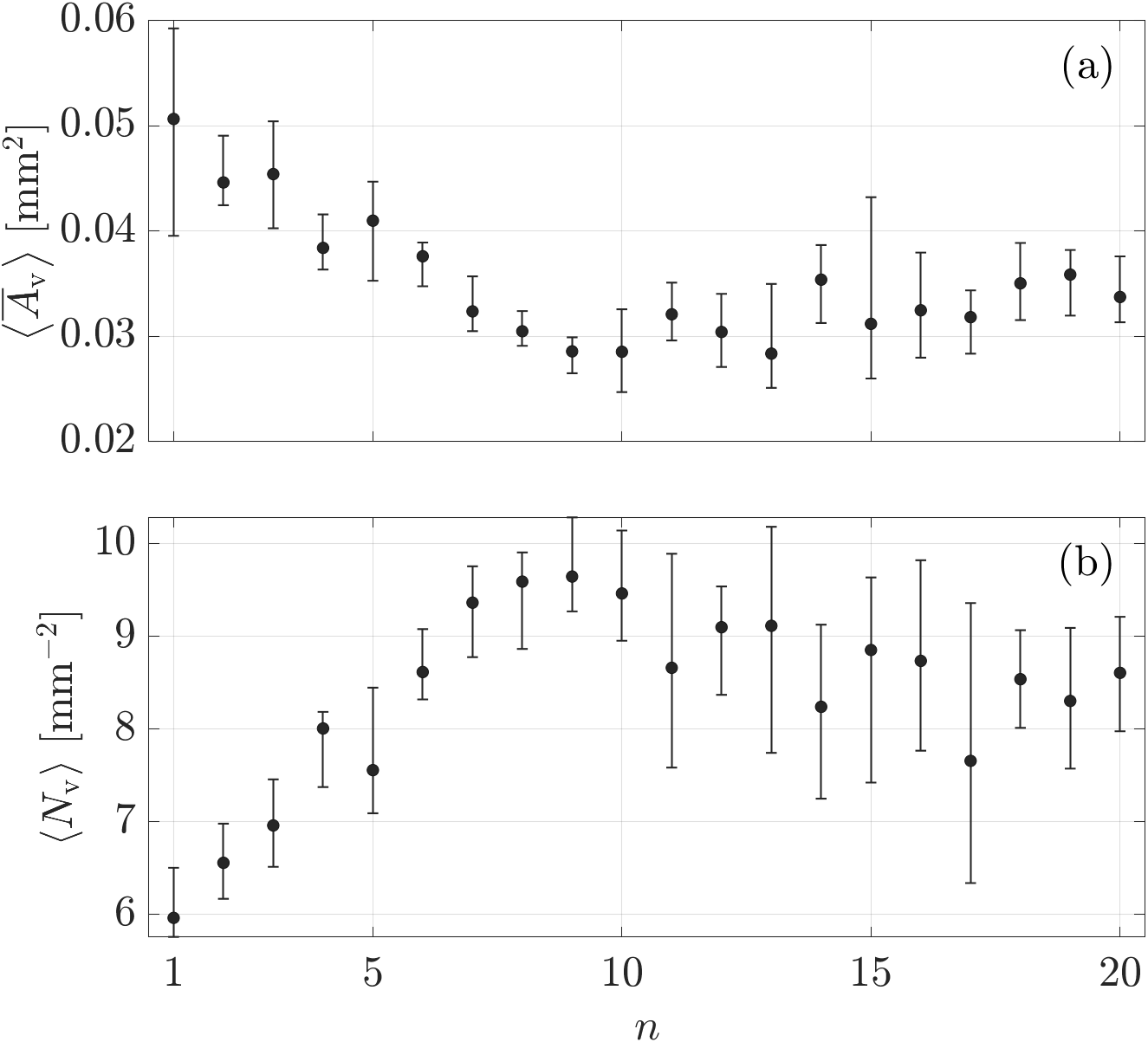}
\caption{ (a) Mean void area $ \left\langle \overline{A}_\mathrm{v} \right\rangle$ and (b) void number density $ \left\langle{N}_\mathrm{v} \right\rangle$; 
Bars indicate the minimum and maximum  values observed of $\overline{A}_\mathrm{v}(t)  $ and ${N}_\mathrm{v}(t)$ over this time interval.}
\label{fig:voids}

\end{figure}

While the shear-rate maps characterize deformation within the confined raft, they provide limited information on how the particles reorganize once they leave the confined domain. To investigate possible changes in this behavior, we now examine the void structures developing in the escaped assembly.

Following the procedure described in Sec.~\ref{sec:voids}, we estimate for each run, both the average   void size, $\overline{A}_\mathrm{v}$, and their average number per unit surface, $N_\mathrm{v}$.

The evolution of $\overline{A}_\mathrm{v}$ and  $\overline{N}_\mathrm{v}$ with run number reveals two distinct regimes. During the first stage, extending approximately from runs 1 to 8, the mean void area decreases while the number of voids increases, both significantly. This simultaneous evolution indicates a profound modification of the relaxation process. Rather than occurring through a limited number of large fractures, deformation becomes distributed among a larger number of smaller rearrangement events. The escaped assembly therefore evolves from a relatively rigid and brittle structure toward a more deformable one capable of  many local rearrangements.

For run numbers larger than approximately 8, the mean void area and void number density evolve much more weakly with either a stabilization or a light reverse trend. The void number density also shows larger differences between its minimum and maximum values. Based on these void statistics alone, it is not possible to determine whether the raft has reached a final aged state or whether aging continues without producing  strong effects on the void observables.

The second interpretation is supported by the velocity-profile and shear-rate measurements reported above. Both continue to evolve beyond $n \approx 10$, indicating that the raft mobility and deformability are still changing. The apparent stabilization of the void statistics should therefore be interpreted cautiously: it more likely reflects a reduced sensitivity of these metrics to the ongoing evolution than a definitive end of the aging process.

\subsubsection{Effects on static raft properties}

\begin{figure*}[t]
\centering
\includegraphics[width=0.8\textwidth]{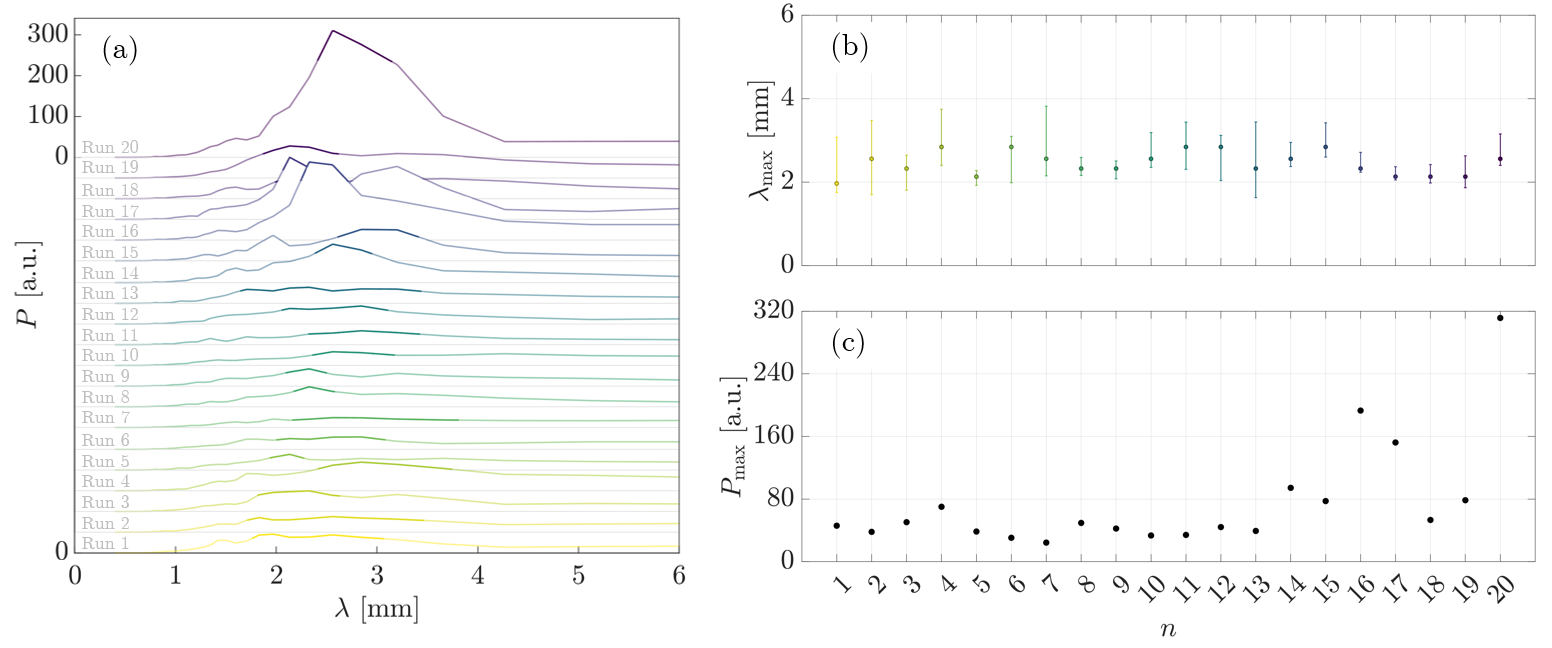}
\caption{(a) Power spectral density $P(\lambda)$ for each run with arbitrary vertical offset for clarity. (b) Characteristic wavelength, $\lambda_\mathrm{max}$,  as a function of run number with bars marking $\lambda_\mathrm{small}$ and $\lambda_\mathrm{large}$, the wavelengths corresponding to \qty{75}{\percent} of the maximum intensity. (c) Maximum spectral density $P_\mathrm{max}$ as a function of run number $n$.}
\label{fig:lambda}
\end{figure*}

This section is dedicated to assessing if the aging observed during dynamic regimes can be measured under static conditions. One classical way to characterize macroscopic mechanical raft properties is to deduce from their wrinkling behavior, and more precisely from  a typical wavelength $ \lambda$, an elastic bending modulus scaling as $B \propto \lambda^{4}$~\cite{vella2004}.  Wrinkles developing within the compressed raft can be visualized with a grazing illumination. Applying to these pictures the method described in Sec.~\ref{sec:wrinkles}, we report in Fig.~\ref{fig:lambda}\,a the power spectral density obtained on the compressed raft before each run. This analysis shows if a wavelength emerges, at which value, and how clearly. Due to the relation between $B$ and $\lambda$ the evolution of the latter provides information about the one of the former. First observations indicate that for each run, a single peak can be identified. While its position remains comparable its magnitude significantly increases with the run number.

These evolutions are quantified in Fig.~\ref{fig:lambda}\,b and c, which represent the main wavelength, $\lambda_\mathrm{max}$ and the peak intensity $P_\mathrm{max}$ as a function of the run number, respectively. Neither the value of $\lambda_\mathrm{max}$ nor the peak width estimated as $\lambda_\mathrm{large}-\lambda_\mathrm{small}$ exhibit a significant evolution over the 20 runs. This means that within the accuracy of the present measurement, the characteristic wavelength of the wrinkling pattern remains unchanged. Thus and since $B \propto \lambda^{4}$, this finding suggests that the effective bending stiffness measured at the raft scale does not vary strongly over the run number. 


In contrast, the peak intensity shows an increasing tendency especially in the last runs. This evolution is also directly visible on the raft images (not shown), where the wrinkling pattern becomes progressively more pronounced and easier to identify with less wrinkle branching for example. From this we conclude that despite similar characteristic wavelength, changes occur within the raft which shows an increased ability to develop this well-defined  pattern coherently at its own scale.

A possible - yet tentative - interpretation is that strong force-chain structures present in young rafts fragment the assembly into several weakly connected domains, leading to branched or interrupted wrinkles. As aging progresses and deformation becomes more distributed, these constraints can progressively relax and allow the emergence of longer and more coherent patterns. 

\subsection{\label{sec:multiscale}Multiscale interpretation  }

\subsubsection{Emerging macroscopic interpretation}
The results reported above reveal a robust and reproducible aging behavior, observed over three independent series of 20 to 30 consecutive runs. 

The different observables consistently indicate that repeated cycling progressively redistributes deformation within the raft. At the macroscopic scale, larger escaped and unjammed areas reveal an increasing ability of the raft to relax. At the mesoscopic scale, higher particle velocities, broader shear zones, the transition from plug-like to parabolic velocity profiles, and the evolution of the escaped assembly toward numerous smaller voids all point to enhanced relative particle motion. Even under static conditions, aging leaves a measurable signature through increasingly coherent wrinkle patterns despite an almost unchanged characteristic wavelength.

Since the global loading conditions remain essentially unchanged, the observed evolution cannot be attributed to variations in the applied load but must result from modifications of stress transmission within the raft. In granular media, stresses are transmitted through heterogeneous force-chain networks carrying a large fraction of the load \cite{cates1999, majmudar2005}. The importance of such structures in particle rafts has previously been evidenced by the strong dependence of unjamming on the constriction location and by arch-like screening effects \cite{plohl2022}. Within this framework, the present results indicate that repeated relaxation progressively modifies the way stresses are transmitted through the raft. Whether aging primarily redistributes the force-chain network itself or instead alters the efficiency of individual particle-particle contacts to transmit stresses cannot be determined from the present measurements. Interestingly, this evolution closely resembles that previously reported for rafts prepared under increasing mixing intensity \cite{planchette2025}, where stronger mixing produced a more homogeneous and cooperative mechanical response. Following the interpretation proposed in that work, the repeated passage through the constriction appears to progressively transform the raft from a more \textit{tempered} toward a more \textit{annealed} state. In this phenomenological picture, repeated rearrangements gradually relax local constraints, allowing the raft to access less frustrated configurations characterized by more homogeneous stress transmission and more cooperative deformation.

This emerging picture is also consistent with recent work on rolling liquid marbles, where the macroscopic response of a particulate interface was shown to be governed by particle-scale friction and shear-induced rearrangements rather than by the properties of an ideal elastic shell \cite{takai2026}.

\subsubsection{\label{sec:microorigin} Toward the microscopic origin}

\begin{figure}[h]
\centering
\includegraphics[width=0.98\linewidth]{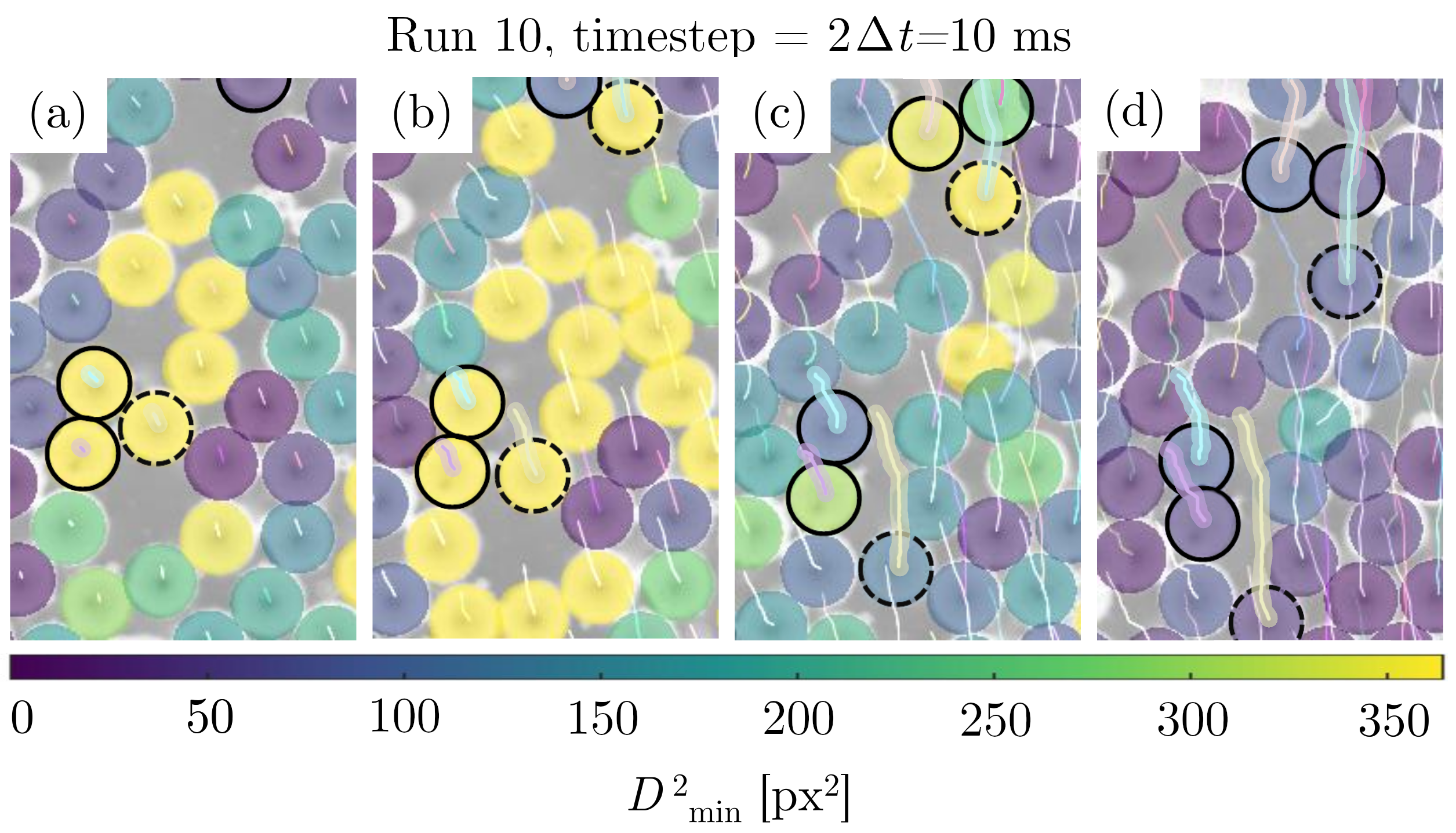}
\caption{Particle tracking in the high-shear region ith time step of $\displaystyle{2\,\Delta t = \SI{10}{\milli\second}}$ (run 10). (a–-d) Rearrangement sequence with particle colors providing the magnitude of $D^2_{\min}$ and continuous lines their respective trajectories. Solid/dashed circles mark particles that keep/loose contact with their neighbor.}
\label{fig:movement}
\end{figure}

The observables reported above all point to a
progressive redistribution of deformation, but none of them reveals the
elementary events responsible for it. To identify the underlying microscopic mechanisms, we examine the particle rearrangements occurring during relaxation.

Figure~\ref{fig:movement}  shows a representative sequence of frames separated by $\displaystyle{2\,\Delta t = \SI{10}{\milli\second}}$ recorded upstream of the constriction marked by a dashed rectangle in Fig.~\ref{fig:meso_variables}. Particles are colored according to their non-affine displacement $D^2_{\min}$, which identifies local rearrangements relative to their neighbors \cite{falk1998}. Particles moving approximately affinely with their local neighborhood appear dark, whereas bright colors indicate large non-affine displacements associated with local rolling, sliding and neighbor exchanges. Together with the overlaid trajectories (lines), the sequence demonstrates that the converging flow is accompanied by continuous particle rearrangements within the confined region, rather than by rigid-body translation of the assembly.

The present measurements do not, however, allow these rearrangements to be quantified as a function of run number. Because the observation window is fixed, it intersects the high-shear region only after this region broadens during aging (Fig.~\ref{fig:shear1}). Moreover, aging itself modifies the local flow, making quantitative comparisons between runs difficult.

The observations in Fig.~\ref{fig:movement} demonstrate that particles undergo repeated rolling, sliding and neighbor exchanges while flowing toward the constriction. These elementary rearrangements therefore constitute a possible microscopic mechanism for the development of the macroscopic aging reported above. However, for the aging effects to accumulate over successive runs, they must progressively modify microscopic properties that persist after relaxation. 

One plausible candidate is the contact-line configuration surrounding each particle. Interface distortions associated with pinned or undulated contact lines mediate the magnitude and directionality of lateral capillary interactions between neighboring particles \cite{danov2005,lewandowski2009,liu2018}. Even subtle contact-line modifications may progressively alter cohesion and stress transmission within the raft. This interpretation is consistent with Fournier and Galatola \cite{fournier2002}, who showed that pinned contact lines promote anisotropic interactions and jamming in interfacial particle assemblies.

Importantly, contact lines themselves are known to occupy metastable states and exhibit slow relaxation, memory and physical aging \cite{kaz2012,colosqui2013,wang2016,xue2014}. Repeated rolling and sliding at the constriction could therefore progressively modify the contact-line configuration, for example through successive pinning and depinning events, and consequently alter the capillary interactions involved in stress transmission.

Although the present experiments cannot directly probe these contact-line dynamics, interferometric techniques currently under development in our laboratory offer a promising route to test this hypothesis and investigate the microscopic origin of aging.

\section{Conclusion}
The present study demonstrates that particle rafts subjected to repeated relaxation through a constriction undergo a reproducible and progressive aging process. Similar trends are observed for both front and back compressed rafts, although their magnitude and apparent saturation behavior differ probably due to opposite force-chain direction. Control experiments further show that cyclic compression alone cannot account for the observed evolution, leaving the repeated passage of particles through the constriction being the primary source of aging.

Aging is found to be a continuous process rather than a sudden transition between two states. Across all observables investigated, repeated recycling progressively increases particle mobility and modifies the collective mechanical raft response. The relaxation process evolves from a localized and brittle-like behavior characterized by plug-like flow, concentrated deformation and a limited number of large rearrangements, toward a more cooperative response involving broader shear zones, enhanced wall slip, more distributed rearrangements, and increasingly coherent wrinkling patterns in the final static state.

The microscopic origin of this evolution remains an open question. The present observations suggest that repeated rolling and sliding at the constriction progressively modify particle-particle contacts, possibly through an evolution of the contact-line configuration responsible for lateral capillary interactions.  Establishing this link will require measurements at the scale of individual particles. Interferometric techniques currently under development provide a promising route.

Beyond its fundamental interest, the constriction geometry introduced here may provide a simple and sensitive tool to probe the aging behavior of particle rafts. Future studies could exploit this approach to investigate how particle size, shape, roughness, wettability, friction, or polydispersity influence the rate and magnitude of the aging and which macroscopic effects can be expected. More generally, combining such macroscopic aging measurements with particle-scale observations may ultimately establish direct links between the evolution of contact-line configurations and the collective mechanical response of particle-laden interfaces.

\begin{acknowledgments}
This research was funded in whole or in part by the Austrian Science Fund (FWF) [Grant DOI 10.55776/P33514].  KS acknowledges the support by a fellowship of the German Academic Exchange Service (DAAD).
\end{acknowledgments}

\appendix


\section{\label{app:subsec1}Reproducibility of the experiments}

\begin{figure}[h]
\centering
\includegraphics[width=0.95\linewidth]{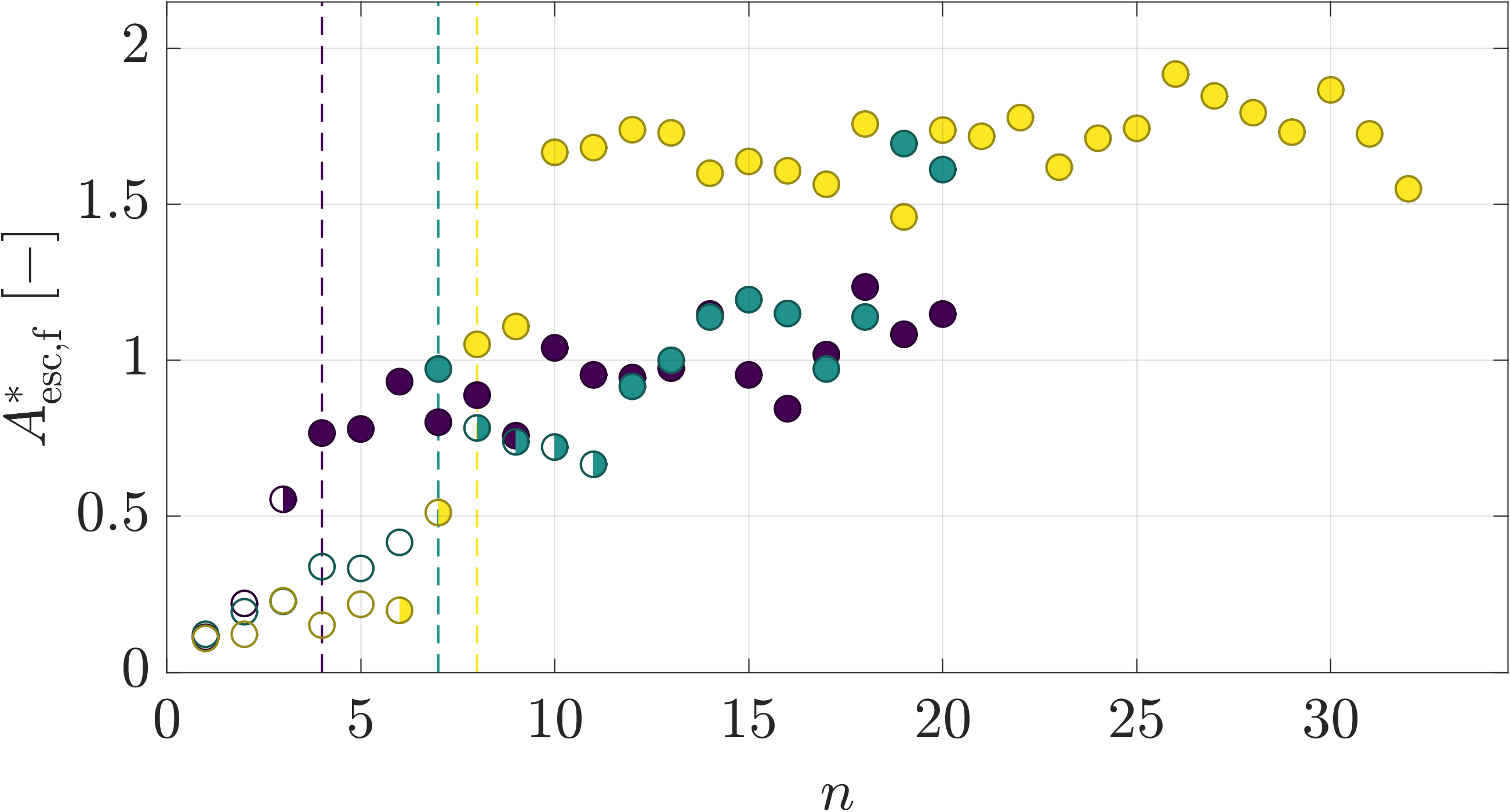}
\caption{Final normalized escaped area $A_{\mathrm{esc,\,f}}^{*}$ as a function of run number $n$ for three independent series distinguished by color (front compression). Empty/half-filled/filled symbols indicate unjamming regions reaching none/one/both lateral walls.}
\label{fig:sizes}
\end{figure}

Using front compression, three sequences of 20 or 30 runs have been conducted and compared by plotting the final normalized escaped area as a function of the run number, see Fig.\ref{fig:sizes}. As it can be seen and despite moderate series to series deviations that cannot be avoided given the granular nature of the system, similar aging is observed on each series. The first extension of the unjammed domain to both lateral walls occur at comparable run number in each series, i.e. run numbers 4, 7 and 8, see dashed lines.

\section{\label{app:subsec2}Validation of the PIV}

\begin{figure*}[t]
\centering
\includegraphics[width=0.95\linewidth]{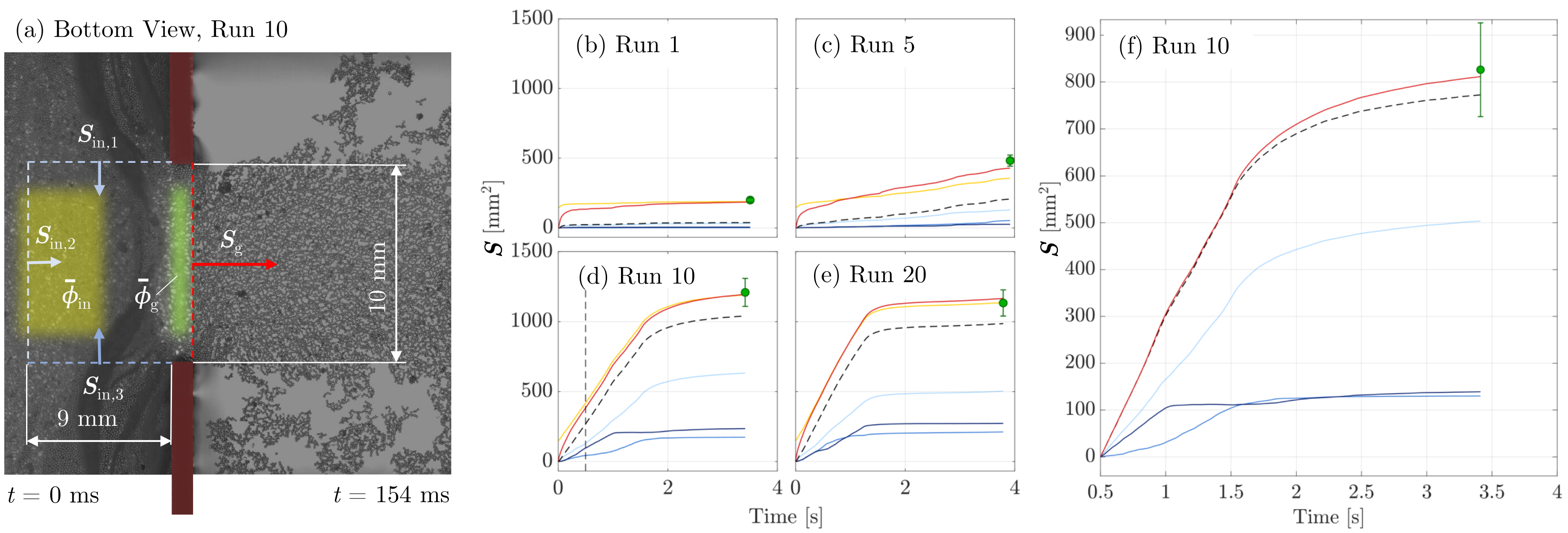}
\caption{(a) Bottom view of run~10: the constriction is marked by the red dashed line with corresponding particle density $\overline{\phi}_\mathrm{g}$ calculated over the green area providing $\mathcal{S}_\mathrm{g}$ (red arrow). The inner domain is contained by the three blue dashed lines $\mathcal{S}_{\mathrm{in,\,}i}$ ($i=1..3$) with the yellow area used to calculate the local particle density $\overline{\phi}_\mathrm{in}$. (b--e) Runs 1, 5, 10, and 20, with black dashed curve: $\mathcal{S}_\mathrm{in,\,tot}$; yellow curve: $\mathcal{S}_\mathrm{in,\,tot}$ with an arbitrary offset of $\approx\SI{150}{\milli\metre\squared}$ to correct for uncaptured folds at the beginning of the relaxation; green symbol: independently measured escaped area $A_\mathrm{esc,\,4s}$ at \SI{4}{\second} with its uncertainty. (f) Run~10: similar results obtained by shifting the time origin of $t=\SI{0.5}{\second}$ to eliminate the fold-induced offset.}
\label{fig:validation_PIV}
\end{figure*}
To validate the accuracy of our PIV analysis, we compared three different metrics describing the quantity of escaped particles over time to each other.

The first metric is built on $\overline{Q}_\mathrm{g} (t, n) =\overline{\phi}_\mathrm{g} (t, n) \,\overline{u}_x(t,n)$ the spatially averaged particle flux obtained at the constriction, which is calculated according to the method described in Sec.~\ref{sec:piv_gate}. Practically, this flux  is multiplied by the constriction width  and integrated over time providing the cumulative amount of escaped particles, measured by the area they collectively occupy, i.e. by $\displaystyle{\mathcal{S}_\mathrm{g}=\int\limits_0^t w \,\overline{Q}_\mathrm{g}\,\mathrm{d}t}$. 

The second metric we use is based on the flux of particles calculated upstream of the constriction using the pale blue dashed lines and yellow area shown in Fig.~\ref{fig:validation_PIV}. Validating the velocity field obtained in this region is of great importance since it is further used to build shear rate maps. Similarly to the analysis performed at the constriction, the velocity field obtained in the confined domain is projected perpendicularly to the three pale blue dashed lines enclosing this sub-domain. They are spatially averaged along each line $i$ and multiplied by each line length to give a two dimensional flux $\overline{V}_{\mathrm{in,\,}i}$  in $\mathrm{m}^2/\mathrm{s}$. The particle flux is then obtained by multiplying each of  these three fluxes ($i=1..3$) with $\overline{\phi}_{\mathrm{in}}$ the spatially averaged particle  packing fraction obtained over the yellow area, which leads to  $\displaystyle{\mathcal{S}_{\mathrm{in,\,}i}= \overline{V}_{\mathrm{in,\,}i}\,\overline{\phi}_{\mathrm{in}}}$. Note that $\overline{\phi}_{\mathrm{in}}$ is computed using a thresholding function as described for the constriction area in Sec.~\ref{sec:piv_gate}. Finally and similarly to $\mathcal{S}_\mathrm{g}$, the constriction metric, this inner flux is integrated over time providing the cumulative amount of particle passing the dashed lines. For each line, we refer to it as to $\displaystyle{\mathcal{S}_{\mathrm{in,\,}i}=\int\limits_0^t \overline{V}_{\mathrm{in,\,}i}\, \overline{\phi}_{\mathrm{in}}\, \mathrm{d}t}$.  The sum over the three lines, i.e. $\displaystyle{\mathcal{S}_\mathrm{in, \,tot}=\sum \limits_{i=1}^{3} {\mathcal{S}_{\mathrm{in,\,}i}}}$, therefore represents the surface occupied by the particles that crossed the lines drawn in the confined domain.

The last and third metric to which to compare with is the final escaped area $A_\mathrm{esc,\,4s}$. 

These three metrics are validated against each other for four representative runs, namely runs 1, 5, 10 and 20, see Fig.~\ref{fig:validation_PIV}\,a--d.
The green symbols represent the  estimates of the final escaped areas with the error bars indicating the uncertainty of these measurements. The latter are caused principally by imperfect thresholding and/or resolution limit for detecting small gaps between particles. The red curves correspond to $\mathcal{S}_\mathrm{g}$ i.e. the surface occupied by the particles that flew through the  constriction (red dashed line in Fig.~\ref{fig:validation_PIV}). The dashed black line shows $\mathcal{S}_\mathrm{in}$, i.e. the surface occupied by the particles that  passed across the dashed lines located in the confined domain. It corresponds to the sum of   the three blue curves, each giving the contribution of one  of these three lines. 

We observe that the final values provided by the red lines are in very good agreement with the green symbols validating the PIV performed at the constriction.
Comparing now the black dashed line to the red one, we see similar evolution with a slight discrepancy corresponding to an almost constant offset. The latter can be well explained by the fact that at early times, folds remain present in the confined area, see Fig.~\ref{fig:validation_PIV}. These are not accounting for by the mean packing fraction $\overline{\phi}_{\mathrm{in}}$. Adding a typical surface of approximately \SI{150}{\milli\metre\squared} to $\mathcal{S}_\mathrm{in, \,tot}$ provides the yellow curves that are in very good agreement with the red ones.

This correction can be done in a more quantitative manner by shifting the time origin from each the metrics are calculated.  Fixing this new time origin as the instant marked by  the vertical dashed line of Fig.~\ref{fig:validation_PIV}\,d, we obtain the curves of Fig.~\ref{fig:validation_PIV}\,f. Discrepancies are smaller than measurement uncertainties caused by limited spatial resolution and imperfect illumination. This comparison finally allows us to validate the velocity fields obtained at the constriction and in the confined domain, further validating velocity profiles, shear rate maps and any other  quantity derived from the PIV.


\bibliography{apssamp}

@article{plohl2022,
  title = {Unjamming strongly compressed rafts: Effects of the compression direction},
  author = {Plohl, Gregor and Jannet, Mathieu and Planchette, Carole},
  journal = {Phys. Rev. E},
  volume = {106},
  issue = {3},
  pages = {034903},
  numpages = {14},
  year = {2022},
  month = {Sep},
  publisher = {American Physical Society},
  doi = {10.1103/PhysRevE.106.034903},
  url = {https://link.aps.org/doi/10.1103/PhysRevE.106.034903}
}

@Article{planchette2025,
author ="Planchette, Carole and Plohl, Gregor",
title  ="Unjamming of particle–laden interfaces: effects of geometry and history",
journal  ="Soft Matter",
year  ="2025",
volume  ="21",
issue  ="9",
pages  ="1718-1730",
publisher  ="The Royal Society of Chemistry",
doi  ="10.1039/D4SM01440E",
url  ="http://dx.doi.org/10.1039/D4SM01440E"}

@article{protiere2017,
  title = {Sinking a Granular Raft},
  author = {Proti\`ere, Suzie and Josserand, Christophe and Aristoff, Jeffrey M. and Stone, Howard A. and Abkarian, Manouk},
  journal = {Phys. Rev. Lett.},
  volume = {118},
  issue = {10},
  pages = {108001},
  numpages = {5},
  year = {2017},
  month = {Mar},
  publisher = {American Physical Society},
  doi = {10.1103/PhysRevLett.118.108001},
  url = {https://link.aps.org/doi/10.1103/PhysRevLett.118.108001}
}

@article{vella2004,
doi = {10.1209/epl/i2004-10202-x},
url = {https://doi.org/10.1209/epl/i2004-10202-x},
year = {2004},
month = {oct},
publisher = {},
volume = {68},
number = {2},
pages = {212},
author = {D. Vella and P. Aussillous and L. Mahadevan},
title = {Elasticity of an interfacial particle raft},
journal = {Europhysics Letters}
}

@Article{planchette2012,
author ="Planchette, Carole and Lorenceau, Elise and Biance, Anne-Laure",
title  ="Surface wave on a particle raft",
journal  ="Soft Matter",
year  ="2012",
volume  ="8",
issue  ="8",
pages  ="2444-2451",
publisher  ="The Royal Society of Chemistry",
doi  ="10.1039/C2SM06859A",
url  ="http://dx.doi.org/10.1039/C2SM06859A"
}

@article{petit2016,
  title = {Bending modulus of bidisperse particle rafts: Local and collective contributions},
  author = {Petit, Pauline and Biance, Anne-Laure and Lorenceau, Elise and Planchette, Carole},
  journal = {Phys. Rev. E},
  volume = {93},
  issue = {4},
  pages = {042802},
  numpages = {7},
  year = {2016},
  month = {Apr},
  publisher = {American Physical Society},
  doi = {10.1103/PhysRevE.93.042802},
  url = {https://link.aps.org/doi/10.1103/PhysRevE.93.042802}
}

@Article{planchette2018,
author ="Planchette, Carole and Lorenceau, Elise and Biance, Anne-Laure",
title  ="Rupture of granular rafts: effects of particle mobility and polydispersity",
journal  ="Soft Matter",
year  ="2018",
volume  ="14",
issue  ="31",
pages  ="6419-6430",
publisher  ="The Royal Society of Chemistry",
doi  ="10.1039/C8SM00653A",
url  ="http://dx.doi.org/10.1039/C8SM00653A"
}

@article{thielicke2014,
 author = {Thielicke, William and Stamhuis, Eize J.},
 doi = {10.5334/jors.bl},
 journal = {Journal of Open Research Software},
 month = {Oct},
 title = {PIVlab – Towards User-friendly, Affordable and Accurate Digital Particle Image Velocimetry in MATLAB},
 year = {2014}
}

@article{thielicke2021,
 author = {Thielicke, William and Sonntag, René},
 doi = {10.5334/jors.334},
 journal = {Journal of Open Research Software},
 month = {May},
 title = {Particle Image Velocimetry for MATLAB: Accuracy and enhanced algorithms in PIVlab},
 year = {2021}
}

@article{taccoen2016,
  title={Probing the mechanical strength of an armored bubble and its implication to particle-stabilized foams},
  author={Taccoen, Nicolas and Lequeux, Fran{\c{c}}ois and Gunes, Deniz Z and Baroud, Charles N},
  journal={Physical Review X},
  volume={6},
  number={1},
  pages={011010},
  year={2016},
  publisher={APS}}

@article{saavedra2018,
  title={Progressive friction mobilization and enhanced Janssen's screening in confined granular rafts},
  author={Saavedra V, Oscar and Elettro, Herv{\'e} and Melo, Francisco},
  journal={Physical Review Materials},
  volume={2},
  number={4},
  pages={043603},
  year={2018},
  publisher={APS}
}

@article{arganda2017,
  title={Trainable Weka Segmentation: a machine learning tool for microscopy pixel classification},
  author={Arganda-Carreras, Ignacio and Kaynig, Verena and Rueden, Curtis and Eliceiri, Kevin W and Schindelin, Johannes and Cardona, Albert and Sebastian Seung, H},
  journal={Bioinformatics},
  volume={33},
  number={15},
  pages={2424--2426},
  year={2017},
  publisher={Oxford University Press}
}

@article{milner1989,
  title={Buckling of Langmuir monolayers},
  author={Milner, Scott Thomas and Joanny, JF and Pincus, P},
  journal={EPL (Europhysics Letters)},
  volume={9},
  number={5},
  pages={495--500},
  year={1989}
}

@article{cerda2002,
  title={Wrinkling of an elastic sheet under tension},
  author={Cerda, Enrique and Ravi-Chandar, K and Mahadevan, L},
  journal={Nature},
  volume={419},
  number={6907},
  pages={579--580},
  year={2002},
  publisher={Nature Publishing Group}
}

@article{liu2018,
  title={Capillary assembly of colloids: Interactions on planar and curved interfaces},
  author={Liu, Iris B and Sharifi-Mood, Nima and Stebe, Kathleen J},
  journal={Annual Review of Condensed Matter Physics},
  volume={9},
  pages={283--305},
  year={2018},
  publisher={Annual Reviews}
}

@article{lewandowski2009,
  title={Oriented assembly of anisotropic particles by capillary interactions},
  author={Lewandowski, Eric P and Bernate, Jorge A and Tseng, Alice and Searson, Peter C and Stebe, Kathleeen J},
  journal={Soft Matter},
  volume={5},
  number={4},
  pages={886--890},
  year={2009},
  publisher={Royal Society of Chemistry}
  }

@article{xue2014,
  title={Strongly metastable assemblies of particles at liquid interfaces},
  author={Xue, Nan and Wu, Shuai and Sun, Sijie and Qu{\'e}r{\'e}, David and Zheng, Quanshui},
  journal={Langmuir},
  volume={30},
  number={49},
  pages={14712--14716},
  year={2014},
  publisher={ACS Publications}
}

@article{kaz2012,
  title={Physical ageing of the contact line on colloidal particles at liquid interfaces},
  author={Kaz, David M and McGorty, Ryan and Mani, Madhav and Brenner, Michael P and Manoharan, Vinothan N},
  journal={Nature materials},
  volume={11},
  number={2},
  pages={138--142},
  year={2012},
  publisher={Nature Publishing Group UK London}
}

@article{fournier2002,
  title={Anisotropic capillary interactions and jamming of colloidal particles trapped at a liquid-fluid interface},
  author={Fournier, J-B and Galatola, P},
  journal={Physical Review E},
  volume={65},
  number={3},
  pages={031601},
  year={2002},
  publisher={APS}
}

@article{colosqui2013,
  title={Colloidal Adsorption at Fluid Interfaces: Regime Crossover from Fast Relaxation to Physical Aging},
  author={Colosqui, Carlos E and Morris, Jeffrey F and Koplik, Joel},
  journal={Physical review letters},
  volume={111},
  number={2},
  pages={028302},
  year={2013},
  publisher={APS}}

@article{wang2016,
  title={Contact-line pinning controls how quickly colloidal particles equilibrate with liquid interfaces},
  author={Wang, Anna and McGorty, Ryan and Kaz, David M and Manoharan, Vinothan N},
  journal={Soft Matter},
  volume={12},
  number={43},
  pages={8958--8967},
  year={2016},
  publisher={Royal Society of Chemistry}
}

@article{cicuta2009,
  title={Granular character of particle rafts},
  author={Cicuta, Pietro and Vella, Dominic},
  journal={Physical review letters},
  volume={102},
  number={13},
  pages={138302},
  year={2009},
  publisher={APS}
}

@article{danov2005,
  title={Interactions between particles with an undulated contact line at a fluid interface: Capillary multipoles of arbitrary order},
  author={Danov, Krassimir D and Kralchevsky, Peter A and Naydenov, Boris N and Brenn, G{\"u}nter},
  journal={Journal of colloid and interface science},
  volume={287},
  number={1},
  pages={121--134},
  year={2005},
  publisher={Elsevier}
}

@article{cates1999,
  title={Jamming and static stress transmission in granular materials},
  author={Cates, ME and Wittmer, JP and Bouchaud, J-P and Claudin, P},
  journal={Chaos: An Interdisciplinary Journal of Nonlinear Science},
  volume={9},
  number={3},
  pages={511--522},
  year={1999},
  publisher={American Institute of Physics}
}

@article{majmudar2005,
  title={Contact force measurements and stress-induced anisotropy in granular materials},
  author={Majmudar, Trushant S and Behringer, Robert P},
  journal={nature},
  volume={435},
  number={7045},
  pages={1079--1082},
  year={2005},
  publisher={Nature Publishing Group UK London}
}

@article{falk1998,
  title = {Dynamics of viscoplastic deformation in amorphous solids},
  author = {Falk, M. L. and Langer, J. S.},
  journal = {Phys. Rev. E},
  volume = {57},
  issue = {6},
  pages = {7192--7205},
  numpages = {0},
  year = {1998},
  month = {Jun},
  publisher = {American Physical Society},
  doi = {10.1103/PhysRevE.57.7192},
  url = {https://link.aps.org/doi/10.1103/PhysRevE.57.7192}
}

@article{trottet2025,
author = {Bertil Trottet  and Daisuke Noto  and Douglas J. Jerolmack  and Hugo N. Ulloa },
title = {Sandball genesis from raindrops},
journal = {Proceedings of the National Academy of Sciences},
volume = {122},
number = {52},
pages = {e2519392122},
year = {2025},
doi = {10.1073/pnas.2519392122},
URL = {https://www.pnas.org/doi/abs/10.1073/pnas.2519392122},
eprint = {https://www.pnas.org/doi/pdf/10.1073/pnas.2519392122}
}

@misc{takai2026,
      title={Static friction of liquid marbles}, 
      author={Yui Takai and Kei Mukoyama and Pritam Kumar Roy and Guillaume Lagubeau and David Quéré and Samuel Poincloux and Timothée Mouterde},
      year={2026},
      eprint={2506.20116},
      archivePrefix={arXiv},
      primaryClass={cond-mat.soft},
      url={https://arxiv.org/abs/2506.20116}, 
}

@article{ershov2022,
  author  = {Ershov, Dmitry and Phan, Minh-Son and Pylv{\"a}n{\"a}inen, Joanna W. and Rigaud, St{\'e}phane U. and Le Blanc, Laure and Charles-Orszag, Arnauld and Conway, James R. W. and Laine, Romain F. and Roy, Nathan H. and Bonazzi, Daria and Du{\'e}nas-Decamps, Guillaume and Cur{\'a}, Michael J. and Karsenti, Eric and Lefebvre, Ophelie and Rideau, Alexis and Roux, Antoine and Tinevez, Jean-Yves},
  title   = {{TrackMate} 7: integrating state-of-the-art segmentation algorithms into tracking pipelines},
  journal = {Nature Methods},
  year    = {2022},
  volume  = {19},
  number  = {7},
  pages   = {829--832},
  doi     = {10.1038/s41592-022-01507-1}
}

@article{tinevez2017,
  author  = {Tinevez, Jean-Yves and Perry, Nick and Schindelin, Johannes and Hoopes, Genevieve M. and Reynolds, Gregory D. and Laplantine, Emmanuel and Bednarek, Sebastian Y. and Shorte, Spencer L. and Eliceiri, Kevin W.},
  title   = {{TrackMate}: An open and extensible platform for single-particle tracking},
  journal = {Methods},
  year    = {2017},
  volume  = {115},
  pages   = {80--90},
  doi     = {10.1016/j.ymeth.2016.09.016}
}

\end{document}